\documentclass[%
reprint,
superscriptaddress,
showkeys,
longbibliography,
pre,
amsmath,amssymb,
aps,
twocolumns,
]{revtex4-2}
\usepackage[english]{babel}
\usepackage{mathtools}

\usepackage{graphicx}%
\usepackage{orcidlink}
\usepackage{multirow}%
\usepackage{amsmath,amssymb,amsfonts}%
\usepackage{mathrsfs}%
\usepackage[title]{appendix}%
\usepackage[svgnames]{xcolor}
\usepackage{textcomp}%
\usepackage{booktabs}%
\usepackage{algorithm}%
\usepackage{algorithmicx}%
\usepackage{algpseudocode}%
\usepackage{listings}%
\usepackage{url}
\usepackage{subcaption}
\usepackage{hyperref}

\usepackage[section]{placeins}
\newcommand{\SMIR}{HeSIR}

\newcommand\rev[1]{{\color{black}#1}}

\newcommand{\PoliMi}{Dipartimento di Elettronica, Informazione e Bioingegneria, Politecnico di Milano, Milano, Italy}
\newcommand{\ISC}{Istituto dei Sistemi Complessi, Consiglio Nazionale delle Ricerche, Roma, Italy}
\newcommand{\IAC}{Istituto per le Applicazioni del Calcolo ``Mauro Picone'', Consiglio Nazionale delle Ricerche, Roma, Italy}
\newcommand{\Sapienza}{Dipartimento di Fisica, Università di Roma La Sapienza, Roma, Italy}
\newcommand{\IACSesto}{Istituto per le Applicazioni del Calcolo ``Mauro Picone'', Consiglio Nazionale delle Ricerche, Sesto Fiorentino, Italy}

\begin{document}

\title{Impact of behavioral heterogeneity on epidemic outcome and its mapping into effective network topologies}
%Behavioral heterogeneity in epidemic spreading and a unified mapping to effective network topologies}
%The impact of heterogeneity on epidemics: Insights frLH a modified SIR model}
% abbreviated title (for running head)
\author{Fabio Mazza\,\orcidlink{0000-0001-7338-9908}}
\affiliation{\PoliMi{}}
\author{Gabriele Ricci}
\affiliation{\IAC{}}
\author{Francesca Colaiori}
\affiliation{\ISC{}}
\affiliation{\Sapienza{}}
\author{Stefano Guarino}
\affiliation{\IAC{}}
\author{Sandro Meloni\, \orcidlink{0000-0001-6202-3302}}
\affiliation{\IAC{}}
\affiliation{Institute for Cross-Disciplinary Physics and Complex Systems (IFISC), CSIC-UIB, Palma de Mallorca, Spain}
\affiliation{
Centro Studi e Ricerche ``Enrico Fermi", Roma, Italy} 
% \author{Francesco Pierri}
% \affiliation{\PoliMi{}}
\author{Fabio Saracco\,\orcidlink{0000-0003-0812-5927}}
\affiliation{
Centro Studi e Ricerche ``Enrico Fermi", Roma, Italy}
\affiliation{
IMT School For Advanced Studies Lucca, Lucca, Italy
}
\affiliation{\IACSesto{}}

\begin{abstract}
Human behavior plays a critical role in shaping epidemic trajectories. During health crises, people respond in diverse ways in terms of self-protection and adherence to recommended measures, largely reflecting differences in how individuals assess risk. This behavioral variability induces effective heterogeneity into key epidemic parameters, such as infectivity and susceptibility. We introduce a minimal extension of the susceptible-infected-removed~(SIR) model, denoted HeSIR, that captures these effects through a simple bimodal scheme, where individuals may have higher or lower transmission--related traits. We derive a closed-form expression for the epidemic threshold in terms of the model parameters, and the network's degree distribution and homophily, defined as the tendency of like--risk individuals to preferentially interact.
We identify a resurgence regime just beyond the classical threshold, where the number of infected individuals may initially decline before surging into large-scale transmission.
% This scenario is analytically demonstrated under the homogeneous mixing assumption and 
Through simulations on homogeneous and heterogeneous network topologies we corroborate the analytical results and highlight how variations in susceptibility and infectivity influence the epidemic dynamics.
We further show that, under suitable assumptions, the HeSIR model maps onto a standard SIR process on an appropriately modified contact network, providing a unified interpretation in terms of structural connectivity.  
Our findings quantify the effect of heterogeneous behavioral responses, especially in the presence of homophily, and caution against underestimating epidemic potential in fragmented populations, which may undermine timely containment efforts. 
The results also extend to heterogeneity arising from biological or other non-behavioral sources. 
\end{abstract}

\maketitle              % typeset the title of the contribution

\section{Introduction}\label{sec1}

The presence of heterogeneity at multiple levels—biological, behavioral, and social—plays a crucial role in shaping epidemic spreading dynamics. Early works in network science demonstrated how structural differences, such as degree variability in contact networks, can substantially alter both the epidemic threshold and outbreak sizes~\cite{pastor2015epidemic, newman2003structure, barrat2008dynamical, newman2018networks}. In particular, variations in the number and type of social interactions across individuals significantly influence the overall spreading potential of a pathogen~\cite{pastor2001epidemic, newman2002, volz2008sir}.
Beyond structural heterogeneity, individual-level biological variations—reflected in key epidemic parameters such as susceptibility, infectivity, and infectious period—also play a critical role. Studies have shown that heterogeneity in susceptibility to infection and infectiousness can markedly affect both the epidemic threshold and the final outbreak size~\cite{Lloyd-Smith2005, Cota2021, deArruda2020, zhilan2000, clancy2014parameters, gou2017parameters, rodrigues2009heterogeneity, neri2011heterogeneity, ming2016stochastic, krylova2013, bonaccorsi2016epidemics, montalban2022, Darbon2019, Lloyd2001}. Building on these insights, Miller introduced a modeling framework capable of incorporating general distributions for these parameters and characterized the epidemic threshold as a function of their heterogeneity~\cite{miller2007epidemic}.
In addition to structural and biological heterogeneity, the COVID-19 pandemic has underscored the pivotal role of human behavior in epidemic trajectories. Behavioral variability can arise from economic and social factors, leading to differences in compliance with control measures, as discussed in recent works on epidemic fatigue and adherence~\cite{DeMeijere2021,DiDomenico2021}, and studies on economic disparities in adherence behavior~\cite{valganon2023,Valganon2023-2}. Another significant source of variability stems from information exposure and political affiliation, which have been shown to influence individual risk behaviors during epidemics~\cite{deverna2024modeling}.
Within this context, it is essential to model how different behaviors—potentially driven by information or misinformation—affect disease spread. Since the adoption of such behaviors often depends on homophily among individuals with similar beliefs, investigating the influence of homophily on epidemic dynamics becomes particularly important.

In this work, we develop a generalized SIR model~(HeSIR) that incorporates \rev{individual behavioral heterogeneity in its simplest form through a mixture of two delta functions, corresponding to low-- and high--compliance individuals. We expect this minimal representation to capture the essential behavior of more general bimodal distributions, where the population consists of two groups with distinct risk--related traits.
We thus consider two behavioral classes, low--risk~(L) and high--risk~(H) individuals, where high--risk individuals display enhanced susceptibility and infectivity, quantified by factors  $\alpha_S>1$ and $\alpha_I>1$. These differences in behavioral compliance might reflect exposure to misinformation~\cite{deverna2024modeling, beca-martinez_compliance_2022}.
Modeling compliance as independent heterogeneity in infectivity and susceptibility, rather than as a single “caution” parameter, enables us to capture the asymmetric effects of non--pharmaceutical interventions such as the use of face masks or social distancing. For instance, mask--wearing by a susceptible individual reduces their infection probability (lowering susceptibility), while mask--wearing by an infected individual primarily limits onward transmission (reducing infectivity). Empirical and modeling studies during COVID-19 support this distinction, showing that interventions that mainly protect the adopter (reducing susceptibility) can be more effective in limiting epidemic prevalence than those that primarily protect contacts (reducing infectivity) with comparable efficacy~\cite{pastor2022advantage}.}

Using a heterogeneous mean--field (HMF) approximation, we derive an analytical expression for the epidemic threshold that holds for networks with arbitrary degree distributions and a tunable level of homophily, defined as the tendency of like-risk individuals to preferentially interact. This result generalizes classical threshold conditions derived for homogeneous networks and uncovers non-trivial dependencies on both behavioral heterogeneity and network structure.
To validate our analytical findings, we perform extensive simulations on random networks generated via the Degree-Corrected Stochastic Block Model (DC-SBM)~\cite{holland1983stochastic,newman2003structure}, also considering the special cases of the Stochastic Block Model (SBM), Configuration Model (CM) and Erdős–Rényi (ER) model, obtained in the absence of degree heterogeneity, homophily, or both. In each case, we estimate the epidemic threshold by locating the peak of epidemic variability, a method known to accurately identify critical points in finite-size systems~\cite{castellano2010thresholds}. The excellent agreement between simulation and theory across different network topologies and mixing patterns confirms the robustness of our generalized threshold formula.

Our simple model exhibits a rich dynamical behavior, including the emergence of a ``resurgence zone''\rev{, a phenomenology that we have already described and analyzed in a previous work~\cite{mazza2024impact}.
In a well-mixed population, this corresponds to an interval between two theoretical thresholds, existing only when both $\alpha_S>1$ and $\alpha_I>1$,} in which the infected population initially declines before exploding into a full outbreak.
Interestingly, we find that heterogeneous network structure alone can produce a similar resurgence phenomenon even in the standard SIR limit ($\alpha_S=\alpha_I=1$), suggesting a unifying mechanism through which structural and epidemiological heterogeneity both sustain hidden transmission potential.
%While in the standard SIR model an outbreak requires $\dot I(0)>0$, in the HeSIR model outbreaks can occur even when $\dot I(0)<0$, provided that the effective growth rate  
%\[
%\dot I_\text{eff}(0) = \dot I_L(0) + %\alpha_I\,\dot I_H(0)
%\]
%is positive. 

Finally, we show that, in expectation, the HeSIR model is equivalent to a standard SIR process on a directed SBM with a bimodal out--degree distribution. In this mapping, the parameters $\alpha_S$ and $\alpha_I$ translate into modulations of effective contact frequency rather than per-contact transmission risk, revealing a conceptual duality: heterogeneity in epidemic parameters can be recast as heterogeneity in contact structure without altering threshold conditions.

In this work, we thus provide a unified framework for understanding how multi-level heterogeneity—including biological, structural, and behavioral factors—shapes epidemic thresholds and dynamics, with important implications for risk assessment and intervention design.

\section{Heterogeneous Epidemic Spreading Model}
\label{sec:model}

Let $I_H$, $I_L$, $S_H$, $S_L$, and $R$ denote the fractions of high-risk infected, low-risk infected, high-risk susceptible, low-risk susceptible, and recovered individuals, respectively.
The total population is conserved: $I_H + I_L + S_H + S_L + R = 1$. 
The mean-field equations for the \SMIR{} model read
\begin{equation}
\!\!\left\{\!
\begin{array}{lll}
    \dot{S}_H &= &
    %-\alpha_S \lambda S_H (\alpha_I I_H + I_L) = 
    -\alpha_S \lambda S_H I_{\textrm{eff}} \,,\\
    \dot{S}_L &= &
    %-\lambda S_L (\alpha_I I_H + I_L) = 
    -\lambda S_L I_{\textrm{eff}} \,,\\
    \dot{I}_H &= &
    %\alpha_S \lambda S_H (\alpha_I I_H + I_L) - \gamma I_H = 
    \alpha_S \lambda S_H I_{\textrm{eff}} - \gamma I_H \,,\\
    \dot{I}_L &= &
    %\lambda S_L (\alpha_I I_H + I_L) - \gamma I_L = 
    \lambda S_L I_{\textrm{eff}} - \gamma I_L \,,\\

    \dot{R}   &=& \gamma I\,.
\end{array}
\right.
\label{eq:model}
\end{equation}
%\begin{equation}
%%\!\!\left\{\!
%\!\!\left\{\!
%\begin{array}{l}
%    \dot{I}_H = \alpha_S \lambda S_H (\alpha_I I_H + I_L) - \gamma I_H = \alpha_S \lambda S_H I_{\textrm{eff}} - \gamma I_H \\
%    \dot{I}_L = \lambda S_L (\alpha_I I_H + I_L) - \gamma I_L = \lambda S_L I_{\textrm{eff}} - \gamma I_L \\
%    \dot{S}_H = -\alpha_S \lambda S_H (\alpha_I I_H + I_L) = -\alpha_S \lambda S_H I_{\textrm{eff}} \\
%    \dot{S}_L = -\lambda S_L (\alpha_I I_H + I_L) = -\lambda S_L I_{\textrm{eff}} \\
%    \dot{R}   = \gamma(I_H + I_L)
%\end{array}
%\right.
%\label{eq:model}
%\end{equation}
\rev{$I=I_H+I_L$ denotes the total prevalence of infected individuals, whereas  $I_\text{eff} = I_L+ \alpha_I I_H$ is} the ``effective'' fraction of infected individuals, due to the increased infectivity of the $I_H$ class.
\rev{We set the timescale by fixing the susceptibility of low-risk individuals to 1, and consequently that of high-risk individuals to $\alpha_S$. In the more general case, these susceptibilities could be defined as $\sigma$ and $\alpha_S \sigma$, respectively.}
Individuals in the state $S_L$ thus become infected and flow to $I_L$ at a rate $\lambda I_{\textrm{eff}}$ that is increased by a factor $\alpha_S$ for individuals in the state $S_H$, and 
%$I_{\textrm{eff}} = \alpha_I I_H + I_L$ is the effective density of the infectious population, weighted by relative infectiousness.
$\lambda=\beta \langle k \rangle$ is the transmission rate, given by the individual transmission rate $\beta$ times the average contact capacity of the nodes $\langle k \rangle$.
From both $I_L$ and $I_H$, individuals then flow to $R$ at a rate $\gamma$ by spontaneous recovery.

\subsection{Heterogeneous percolation threshold on a network with homophily}

%\begin{widetext}   
\begin{figure*}%[t!]
    \centering
    \includegraphics[width=\textwidth]{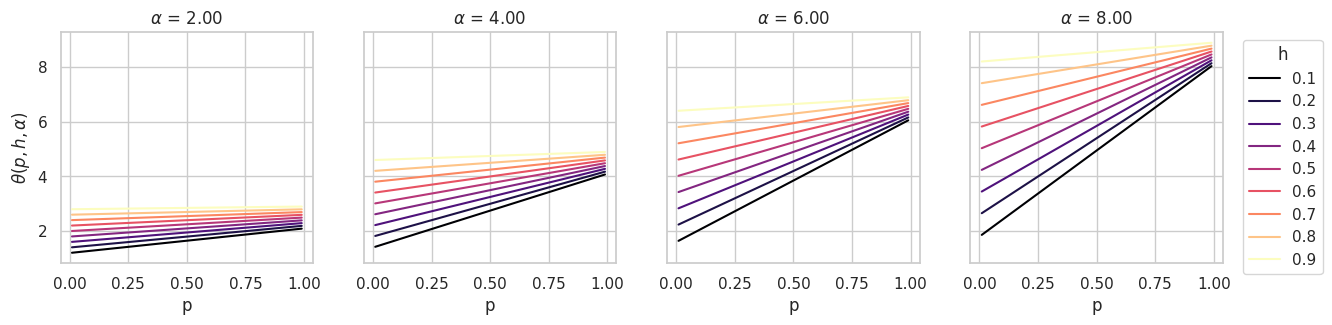}
    \caption{The value of $\theta(p,h,\alpha)$ for different values of $p$, $h$ and $\alpha$. $\theta(p,h,\alpha)$ can be interpreted as the ratio between the threshold in the baseline SIR case and the threshold in the HeSIR model. As expected, $\theta(p,h,\alpha)$ increases with $p$, $h$ and $\alpha$. For large $h$ (resp. $p$), the dependence on $p$ (resp. $h$) weakens.}
    \label{fig:theta}
\end{figure*}
%\end{widetext}

To analyze the model's behavior in a networked population, we consider a degree-corrected SBM~(DC-SBM)~\cite{karrer_stochastic_2011} with two communities: %low-risk ($L$) and high-risk ($H$) individuals, with 
$N_L = (1 - p)N$ low-risk and $N_H = pN$ high-risk individuals.
Let $h \in [0,1]$ be a homophily parameter interpolating between random mixing ($h = 0$) and full assortativity ($h = 1$).
%The limit case $h=1$ corresponds to $O$ and $M$ individuals being segregated into separate connected cLHponents and is considered in the Appendix.
The matrix of conditional probabilities that a randomly chosen neighbor of a node in group $i$ belongs to group $j$ is given by
\begin{equation}
    \begin{pmatrix}
        \pi_{LL} & \pi_{LH} \\
        \pi_{HL} & \pi_{HH}
    \end{pmatrix}
    =
    (1 - h)
    \begin{pmatrix}
        1 - p & p \\
        1 - p & p
    \end{pmatrix}
    + h
    \begin{pmatrix}
        1 & 0 \\
        0 & 1
    \end{pmatrix}\,.
    \label{eq:conditional_edge_probabilities}
\end{equation}
This matrix is defined for $p\in (0, 1)$.
Let $x_L$ and $x_H$ be the probabilities that a vertex of type $L$ or $H$ is not connected to the giant component through a randomly chosen edge. Considering a generic degree distribution, we denote with $q_k$ the corresponding excess degree distribution and with $g_1(x)=\sum_k q_kx^k$ the generating function for the excess degree distribution.
We can determine the size of the giant component by finding the fixed point of the equation
%\begin{widetext}
\begin{equation*}
    \begin{cases}
        x_L = & \pi_{LL}(1 - \phi_{LL} + \phi_{LL} g_1(x_L))\\
              & + \pi_{LH}(1 - \phi_{LH} + \phi_{LH} g_1(x_H)) \,,\\
        x_H = & \pi_{HL}(1 - \phi_{HL} + \phi_{HL} g_1(x_L))\\
              &+ \pi_{HH}(1 - \phi_{HH} + \phi_{HH} g_1(x_H))\,,
    \end{cases}
    \label{eq:percolation_heterogeneous}
\end{equation*}
%\end{widetext}
where the activation probabilities $\phi_{ij}$ (transmission probability across an edge from group $i$ to group $j$) are given by
$\phi_{LL} = 1 - e^{- \beta / \gamma}$, 
$\phi_{LH} = 1 - e^{-\alpha_I  \beta / \gamma} $, $\phi_{HL} = 1 - e^{-\alpha_S  \beta / \gamma}$,
and $\phi_{HH} = 1 - e^{-\alpha_S \alpha_I  \beta / \gamma}$. 
%\begin{equation}
%    \begin{cases}
%        \phi_{LL} = 1 - e^{- \beta / \gamma} \,,\\
%        \phi_{LH} = 1 - e^{-\alpha_I  \beta / \gamma} \,,\\
%        \phi_{HL} = 1 - e^{-\alpha_S  \beta / \gamma} \,.
%    \end{cases}
%    \label{eq:activation_probs}
%\end{equation}
The percolation threshold is found by solving

\begin{align}
    \nonumber
    \left(\pi_{LH}\pi_{HL}\phi_{LH}\phi_{HL}-\pi_{HH}\pi_{LL}\phi_{HH}\phi_{LL}\right) \dot{g}_1^2(1)&\\
    \label{eq:threshold_true}
    + \left(\pi_{LL}\phi_{LL}+\pi_{HH}\phi_{HH}\right)\dot{g}_1(1) -1 &= 0 
\end{align}

% \begin{widetext}
% \begin{align}
%     1 =\frac{\langle k^2\rangle-\langle k \rangle}{\langle k\rangle} \Bigg[ \frac{\langle k^2\rangle-\langle k \rangle}{\langle k\rangle} \left( p(1-p)(1-h)^2\phi_{LH}\phi_{HL} -(p(1-p)(1-h)^2+h)\phi_{LL}\phi_{HH}\right)\nonumber\\
%     +((1-p)(1-h)+h)\phi_{LL}+(p(1-h)+h)\phi_{HH}\Bigg]\,.
%     \label{eq:threshold_true}
% \end{align}
% \end{widetext}
%\begin{widetext}
%\begin{align}    
%%\phi_{LL}\phi_{HH})+h\phi_{LL}\phi_{HH}
%    +\frac{\langle k\rangle}{\langle k^2\rangle-\langle k \rangle}
%    \left[
%    ((1-p)(1-h)+h)\phi_{LL}+(p(1-h)+h)\phi_{HH}
%    \right]-\left[\frac{\langle k\rangle}{\langle k^2\rangle-\langle k \rangle}\right]^2=0
%    \label{eq:threshold_true}
%\end{align}
%\end{widetext}
%\begin{align}     1 &=\frac{\langle k^2\rangle-\langle k \rangle}{\langle k\rangle} \left(p(1-h)\phi_{LH}(1-p)(1-h)\phi_{MO}\frac{\langle k^2\rangle-\langle k \rangle}{\langle k\rangle}\right.\nonumber\\     &+[(1-p)(1-h)+h]\phi_{LL}+[p(1-h)+h]\phi_{HH}\nonumber\\     &\left.-[(1-p)(1-h)+h]\phi_{LL}[p(1-h)+h]\phi_{HH}\frac{\langle k^2\rangle-\langle k \rangle}{\langle k\rangle}\right)    \label{eq:threshold_true} \end{align}
Equation \eqref{eq:threshold_true} admits no closed form solution, but can be solved numerically. 
%we will refer to  solution of \eqref{eq:threshold_true} using numerical methods, such as Newton method.
% Assuming $ \beta / \gamma\ll 1$ and linearizing the $\phi_{ij}$'s, an approximate solution is given by
Further, if we assume $\beta / \gamma\ll 1$, we can linearize \eqref{eq:threshold_true} with respect to $\beta / \gamma$ and ignore the first term of the equation.
Denoting $\alpha=\alpha_I\alpha_S$ and $\theta(p,h,\alpha)=\pi_{LL}+\alpha\pi_{HH}=(1 - h)(1 - p + \alpha p) + h(\alpha + 1)$, we obtain
\begin{equation}\label{eq:threshold_general}
   \left( \frac{\beta}{\gamma} \right)_c = \frac{\langle k \rangle}{\langle k^2 \rangle - \langle k \rangle} \frac{1}{\theta(p,h,\alpha)}
\end{equation}
where we used $\dot{g}_1(1)=\frac{\langle k^2 \rangle - \langle k \rangle}{\langle k \rangle}$.
% \begin{equation}
%     \left( \frac{\beta}{\gamma} \right)_c
%     = \frac{\langle k \rangle}{(\langle k^2 \rangle - \langle k \rangle)\left[ (1 - h)(1 - p + \alpha_S \alpha_I p) + h(\alpha_S \alpha_I + 1) \right]}
%     \label{eq:threshold_general}
% \end{equation}
% \begin{equation}
%     \left( \frac{\beta}{\gamma} \right)_c
%     = \frac{\langle k \rangle}{\langle k^2 \rangle - \langle k \rangle} \frac{1}{\theta(p,h,\alpha)}\,,
%     \label{eq:threshold_general}
% \end{equation}
% where
%  $\theta(p,h,\alpha) = (1 - h)(1 - p + \alpha p) + h(\alpha + 1)$ and $\alpha=\alpha_S\alpha_I$.

\rev{In epidemiological terms, the percolation threshold $\left(\frac{\beta}{\gamma}\right)_c$ corresponds to the critical value of the transmission-to-recovery ratio at which the system undergoes a transition from disease extinction to epidemic outbreak. 
For $\frac{\beta}{\gamma} > \left(\frac{\beta}{\gamma}\right)_c$, the infection can spread within the population, 
whereas for $\frac{\beta}{\gamma} < \left(\frac{\beta}{\gamma}\right)_c$ the outbreak is expected to eventually die out.}

Note that for $\theta=1$ eq.~\eqref{eq:threshold_general} reduces to the epidemic threshold for the standard SIR in a CM.
For small $\beta/\gamma$, the threshold depends on $\alpha_S$ and $\alpha_I$ only through their product $\alpha$.
For a network with no homophily ($h = 0$), $\theta$ simplifies to $\theta(p,0,\alpha) = 1 - p + \alpha p$.
The dependence of $\theta$ on its arguments is shown in Figure~\ref{fig:theta}.
% {\bf questa figura non mi pare molto informativa. che vogliamo dire?}

A detailed analytical derivation of the epidemic threshold is reported in the Appendix.

% %
%     \[
%         \left( \frac{\beta}{\gamma} \right)_c = \frac{\langle k \rangle}{(\langle k^2 \rangle - \langle k \rangle)(1 - p + \alpha_S \alpha_I p)}
%     \]
% for a CM network with no hLHophily ($h = 0$), and to
%     \[
%         \left( \frac{\beta}{\gamma} \right)_c = \frac{1}{(\langle k \rangle - 1)\left[ (1 - h)(1 - p + \alpha_S \alpha_I p) + h(\alpha_S \alpha_I + 1) \right]}
%     \]
% for a SBM network with hLHogeneous degrees ($\langle k^2 \rangle = \langle k \rangle^2$).

%    \item \textbf{Standard SIR} ($\alpha_S = \alpha_I = 1$): recovers the classical result
%    \[
%        \left( \frac{\beta}{\gamma} \right)_c = \frac{1}{(k - 1)}
%    \]

\subsection{Early dynamics and effective threshold}

To analyze the early-time behavior, we assume a small initial infection uniformly distributed across the population: $I_L(0) = (1 - p)\epsilon$, $I_H(0) = p\epsilon$, $S_L(0) \approx 1 - p$, $S_H(0) \approx p$. Setting $R(0) = 0$, we compare the conditions under which
 $\dot{I}(0)=0$ and $\dot{I}_{\textrm{eff}}(0)=0$, that translate in 
\begin{align}
   \rev{\left( \frac{\beta}{\gamma} \right)_{I}} = \frac{1}{\langle k \rangle(1 - p + \alpha_S p)(1 - p + \alpha_I p)} \frac{1}{1 - \epsilon} \label{eq:I_condition} \,,\\
\rev{\left( \frac{\beta}{\gamma} \right)_{I_{\textrm{eff}}}} = \frac{1}{\langle k \rangle(1 - p + \alpha_S \alpha_I p)}  \frac{1}{1 - \epsilon} \,.\label{eq:Ieff_condition}
\end{align} 
 %\begin{align}
%    \dot{I}(0) &= \dot{I}_L(0) + \dot{I}_H(0) = 0 \quad \Rightarrow \quad \left( \frac{\beta}{\gamma} \right) = \frac{1}{\langle k \rangle(1 - p + \alpha_S p)(1 - p + \alpha_I p)} \cdot \frac{1}{1 - \epsilon} \label{eq:I_condition} \\
%    \dot{I}_{\textrm{eff}}(0) &= \alpha_I \dot{I}_H(0) + \dot{I}_L(0) = 0 \quad \Rightarrow \quad \left( \frac{\beta}{\gamma} \right) = \frac{1}{\langle k \rangle(1 - p + \alpha_S \alpha_I p)} \cdot \frac{1}{1 - \epsilon} \label{eq:Ieff_condition}
%\end{align}

In the limit $\epsilon \to 0$ and $\langle k \rangle-1\approx \langle k \rangle$,~\eqref{eq:Ieff_condition} matches our epidemic threshold \eqref{eq:threshold_general} for a homogeneous network without homophily.
On the other hand, again in the limit $\epsilon \to 0$ and $\langle k \rangle-1\approx \langle k \rangle$,~\eqref{eq:I_condition} coincides with the estimate found in~\cite{miller2007epidemic} in the special case of $\langle k\rangle $-regular graphs.
The latter is obtained from bond percolation assuming that all edges have the same average occupation probability 
\begin{equation}
    \phi = (1-p)^2\phi_{LL}+p(1-p)\phi_{LH}+p(1-p)\phi_{HL}+p^2\phi_{HH} \,.
\end{equation}
The two approaches---averaging between each group separately or throughout the network---lead to distinct thresholds only when both $\alpha_S > 1$ and $\alpha_I > 1$.
This underlines the importance of accounting for both heterogeneous susceptibility and infectivity, although \eqref{eq:threshold_general} depends on $\alpha_S$ and $\alpha_I$ only through their product.
The analysis of early dynamics shows that the combined effect of $\alpha_S$ and $\alpha_I$ makes it possible for the total \rev{prevalence} $I(t)$ to initially decrease, while the effective infectious pool $I_{\text{eff}}(t)$ grows enough to eventually trigger a resurgence.

\section{Simulations}
\label{sec:simulation}
\begin{figure*}[ht] %[htbp]
    \centering

    \includegraphics[width=0.9\textwidth]{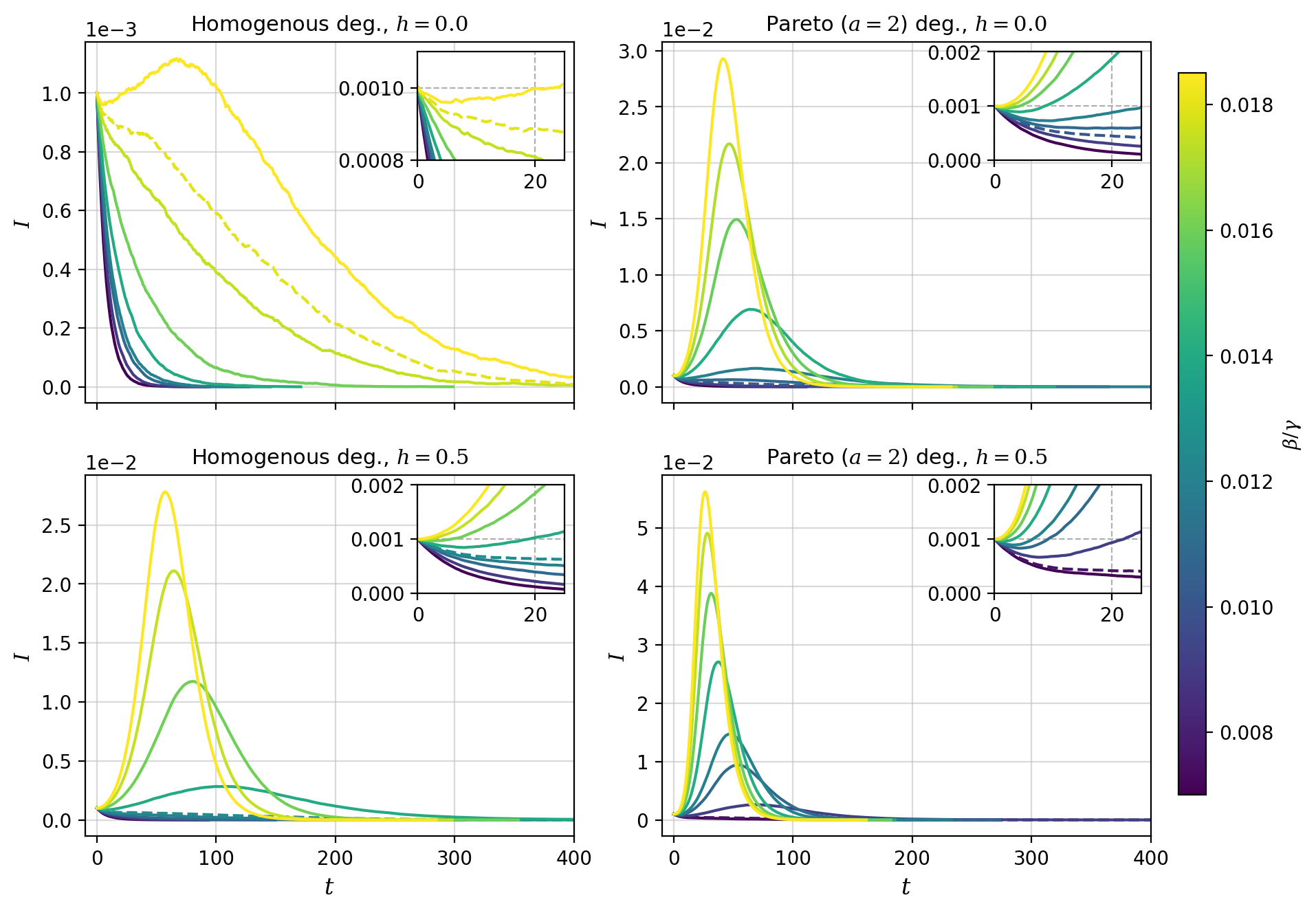}
    \caption{Curves of the mean number of infected under different configurations of degree heterogeneity and homophily, with $p_H=0.4$, $\alpha_I=2$, $\alpha_S=3$.  {Results averaged over $400$ simulations per setting, for networks of $N=5\times 10^4$ nodes with mean degree $\langle k \rangle = 20$, with 50 seed nodes chosen uniformly at random. The dashed lines represent the critical values of $\beta / \gamma$ maximizing the epidemic variability $V_{AR}$. The insets provide a close--up of the outbreak’s early dynamics, highlighting the resurgence zone observed in some super--critical regimes: an initial decline in the infected population followed by renewed growth toward a full outbreak.}
}
    \label{fig:epicurves}
\end{figure*}

To validate the analytical predictions of our model, we performed extensive stochastic simulations of the the HeSIR dynamics on DC-SBM networks. This family of networks generalizes the CM, SBM and ER graphs by tuning the homophily parameter $h$ and the degree distribution.
We simulate HeSIR epidemics with an optimized Gillespie algorithm~\cite{gibson_efficient_2000} in networks with  {$N = 5\times10^4$} nodes and average degree $\langle k \rangle = 20$, with  homogeneous and heterogeneous degree distributions (Pareto degree distributions with exponent $a \in \{2.0, 2.5\}$), and different levels of homophily $h$ and high-risk population fraction $p_H$. 
 {In all simulations, we take $I(0)=10^{-3}$ (i.e. 50 seed nodes when $N = 5\times10^4$, taken uniformly at random) and $R(0)=0$.}

The mixing structure is determined by the edge-frequency matrix:
\begin{equation}
    \begin{pmatrix}
        p_{LL} & p_{LH} \\
        p_{HL} & p_{HH}
    \end{pmatrix}
    \approx \frac{\langle k \rangle}{N}
    \begin{pmatrix}
        \frac{(1 - h)(1 - p) + h}{1 - p} & 1 - h \\
        1 - h & \frac{(1 - h)p + h}{p}
    \end{pmatrix}\,,
    \label{eq:sim_edge_probs}
\end{equation}
which for large $N$ maps \eqref{eq:conditional_edge_probabilities} to the expected frequency of connections between groups for a given $h$ and $p$.

\subsection{Epidemic curves across mixing and heterogeneity}

We analyzed the temporal evolution of the epidemic for different model parameters. Figure~\ref{fig:epicurves} shows representative epidemic curves for the fraction of infected individuals over time, comparing the homogeneous and heterogeneous settings, both with and without homophily.

As seen in the figure, the qualitative shape of the epidemic curve is consistent across all settings. \rev{The presence of degree heterogeneity leads to earlier
% and more intense 
peaks due to superspreading individuals.
By creating clusters of high-risk individuals associated to acute spreading, homophily facilitates the onset and significantly increases peak prevalence.}

\subsection{Epidemic threshold: theory versus simulation}
\begin{figure*}%[ht!]
    \centering
    \begin{subfigure}[b]{0.4\textwidth}
        \centering
        \includegraphics[width=\textwidth]{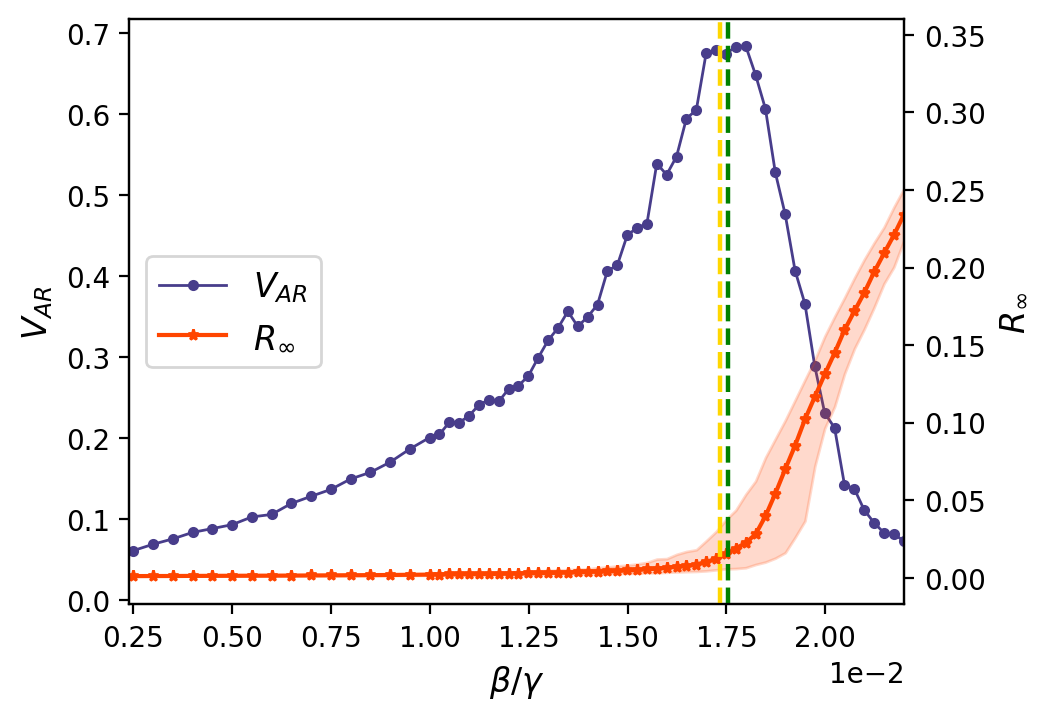}
        %{figures/Grafici_He_SIR/ev_ar_fabio/sbm_h_0_arvar_a_2.0_b_3.0_pm_0.4_fixedsw_Lut_c.png}
        \caption{Homogeneous degrees, $h=0$}
    \end{subfigure}
    \begin{subfigure}[b]{0.4\textwidth}
        \centering
        \includegraphics[width=\textwidth]{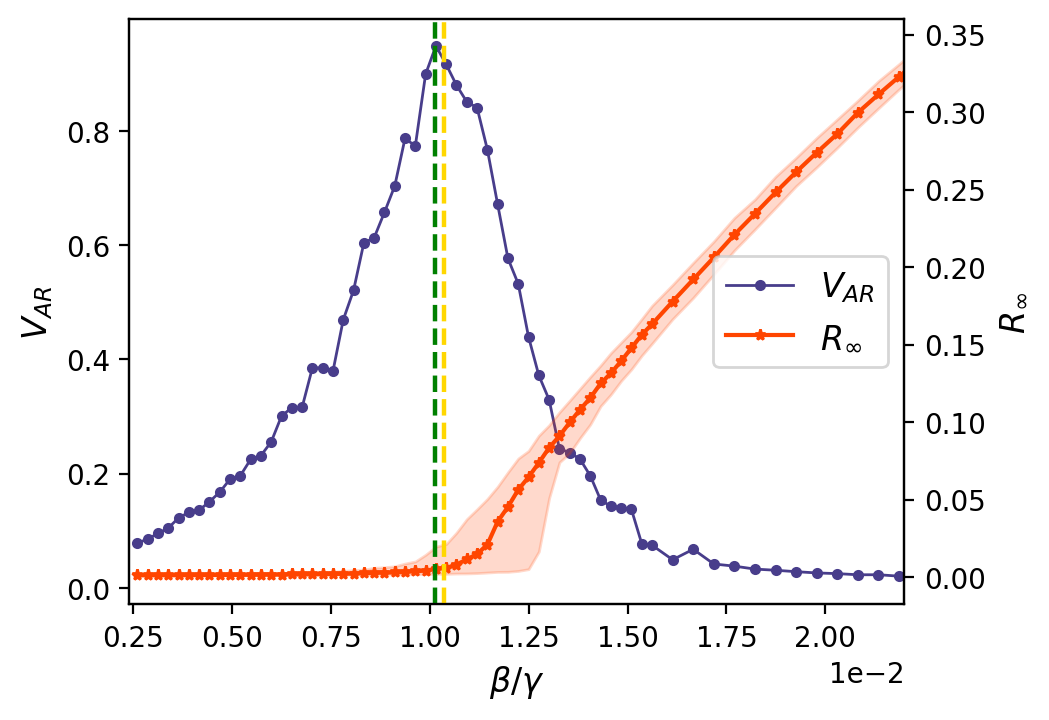}
        \caption{Pareto degrees ($a=2$), $h=0$}
    \end{subfigure}
        \begin{subfigure}[b]{0.4\textwidth}
        \centering
        \includegraphics[width=\textwidth]{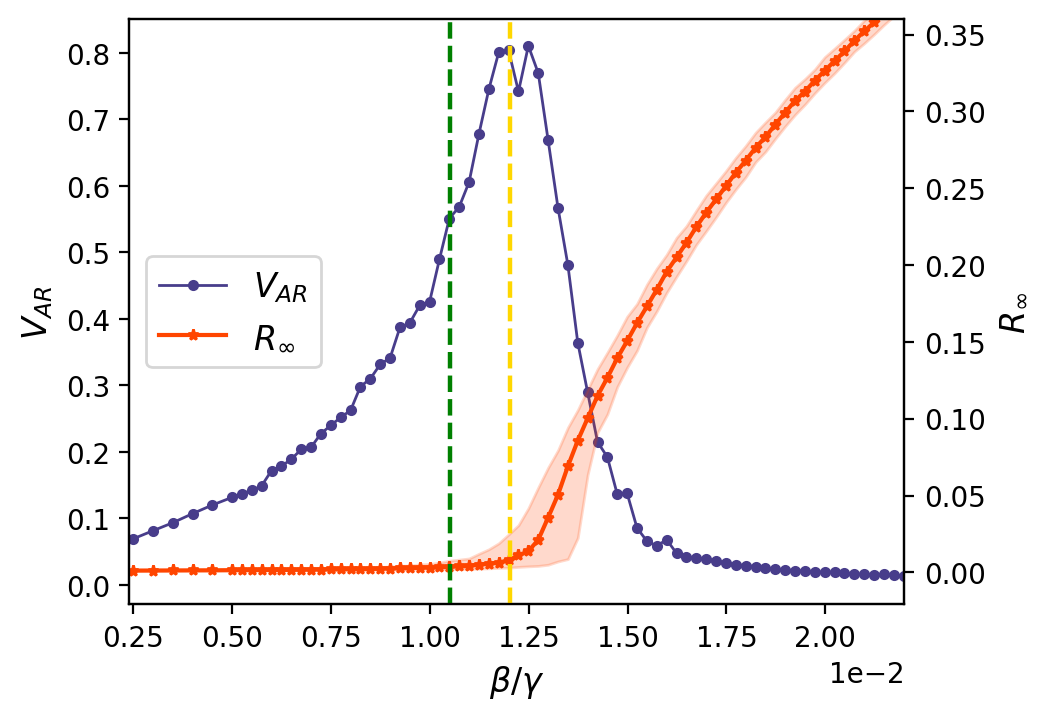}
        \caption{Homogeneous degrees, $h=0.5$}
    \end{subfigure}
    \begin{subfigure}[b]{0.4\textwidth}
        \centering
        \includegraphics[width=\textwidth]{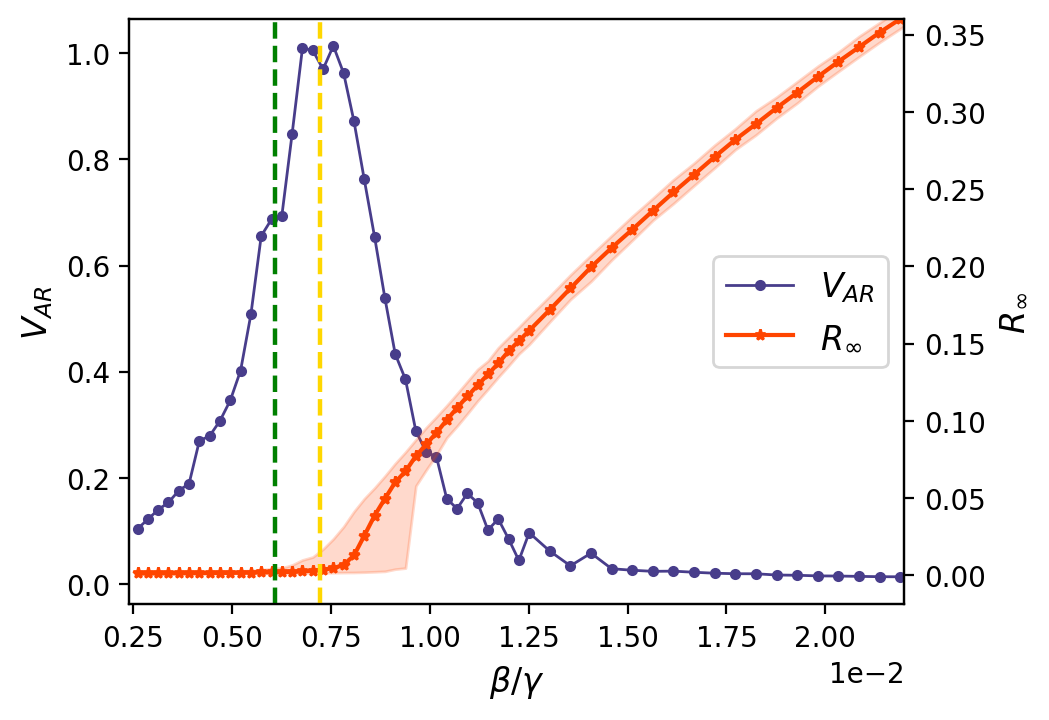}
        \caption{Pareto degrees ($a=2$), $h=0.5$}
    \end{subfigure}
    \caption{Epidemic threshold estimates. Yellow dashed line: analytical threshold, computed numerically; Green dashed line: approximation of the analytical threshold, as given by \eqref{eq:threshold_general}. The $R_\infty$ curves report the median value, with shaded areas showing the central 80\% of simulated outcomes.  {Results are averaged over 400 simulations per setting, for networks of $N = 5 \times 10^4$ nodes and mean degree $\langle k \rangle = 20$, with 50 seed nodes chosen uniformly at random.} Other parameters set as $p_H = 0.4$, $\alpha_I = 2$, and $\alpha_S = 3$. Additional results in the Appendix  {confirm the robustness of the analysis across a range of values for $\alpha_S$, $\alpha_I$, $p_H$, $N$ and $\langle k \rangle$.}}
    \label{fig:thresholds}
\end{figure*}

To assess the accuracy of the analytical epidemic threshold derived in Eq.~\eqref{eq:threshold_general}, we compare three different indicators:
\begin{enumerate}
    \item Theoretical threshold $\left(\beta/\gamma\right)_c$ from heterogeneous percolation.
    \item The value of $\beta/\gamma$ at which the final epidemic size $R_\infty$ exhibits a sharp transition.
    \item The value maximizing the epidemic variability $
        V_\textrm{AR} = \sqrt{\left< R_\infty^2 \right> - \left< R_\infty \right>^2}/\left< R_\infty \right>$  % \FM{ non è il coefficient of variation? }
\end{enumerate}
Figure~\ref{fig:thresholds} illustrates this comparison for various network configurations, with the actual threshold, computed numerically, identified by the yellow dashed line, and the closed-form approximation, given by \eqref{eq:threshold_general}, shown by the green dashed line.

Overall, the numerical estimates for the threshold confirm the analytical predictions for all the cases considered. The position of the phase transition in the attack rate and the peak of $V_\textrm{AR}$ closely match the percolation-based estimates.
%, validating the model’s ability to capture the epidemic onset accurately. 
The small discrepancies observed near the transition point, particularly in highly heterogeneous networks, can be attributed to finite-size effects and stochastic fluctuations near criticality.

%In summary, the simulations confirm that the theoretical threshold derived via heterogeneous percolation is robust across a wide range of network structures and mixing patterns. Both epidemic timing and final size align with analytical expectations.
%, supporting the use of the model for policy analysis and control design.

\section{Mapping the H\lowercase{e}SIR Model onto a SIR model on an Effective Network}

\begin{figure*}[ht!]
    \centering
    \includegraphics[width=\textwidth]{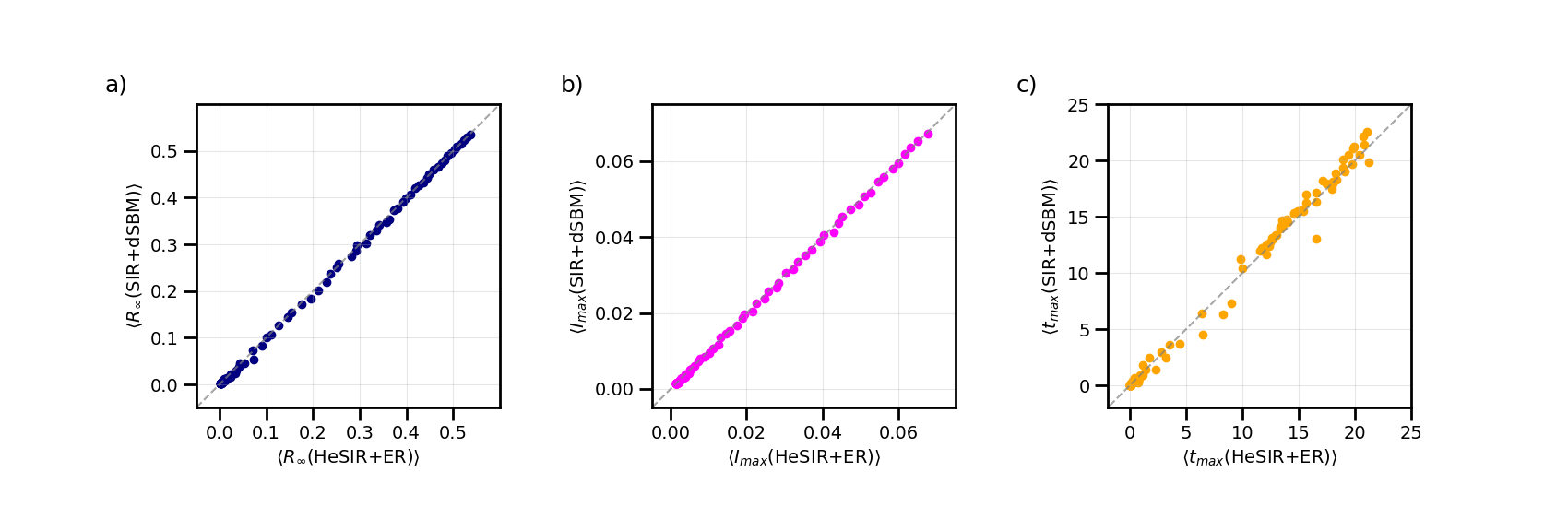}
    \caption{Model equivalence on homogeneous networks. For $\beta/\gamma \in [0.01, 0.04]$, the average values of $R_\infty$ (panel a), \rev{the frequency of infected individuals at the peak} $I_\text{max}$ (panel b), and \rev{the time of the peak} $t_\text{max}$ \rev{(panel c)} are shown for both the HeSIR and the equivalent SIR model, exhibiting perfect agreement. Averages are computed over 200 realizations. Parameters are set to $\alpha_I=2$, $\alpha_S=3$, and $p_H=0.4$.}
    \label{fig:equiv_ER}
\end{figure*}

We now show that the heterogeneous transmission dynamics of the \SMIR{} model can be reformulated as a standard SIR process on an appropriately modified network. Since individuals of type \(H\) differ in both susceptibility (\(\alpha_S \neq 1\)) and infectiousness (\(\alpha_I \neq 1\)), it is natural to ask whether the resulting dynamics could instead be viewed as a homogeneous SIR process on a block-structured contact network in which \(H\) and \(L\) individuals occupy distinct blocks, such that the heterogeneity in susceptibility and infectiousness is effectively transferred to heterogeneity in the connectivity patterns between and within blocks. We demonstrate below that such a representation is indeed possible under certain assumptions.

 {
A related idea---interpreting heterogeneous transmission through an effective directed network under an annealed approximation---appears in work by Clancy~\cite{clancy2018persistence,clancy2018precise}, although in a different setting and for a different purpose: there, the focus is on SIS dynamics and on how heterogeneity in susceptibility and infectiousness influences the persistence time from quasi-stationary endemicity to extinction.
}

%\subsection{Edge-level equivalence}
\subsection{\rev{General intuition}}

Let \(P_i \in \{L, H\}\) denote the class of node \(i\). In the \SMIR{} model, the probability that a susceptible node \(i\) is infected by an infectious node \(j\) is proportional to:
\[
p_{ij}^{\text{\SMIR{}}} \, \beta_{P_jP_i},
\]
where \(p_{ij}^{\text{\SMIR{}}}\) is the connection probability and
\begin{equation}\label{eq:equivalence_0}
\beta_{P_jP_i} = \beta  [1 - (1 - \alpha_S)\delta_{P_i, H}]  [1 - (1 - \alpha_I)\delta_{P_j, H}],
\end{equation}

so that:
\[
\begin{cases}
\beta_{LL} = \beta, \\
\beta_{LH} = \alpha_S \beta, \\
\beta_{HL} = \alpha_I \beta, \\
\beta_{HH} = \alpha_S \alpha_I \beta.
\end{cases}
\]
\rev{We can now recast the dynamics of the HeSIR model as} an equivalent SIR model on an effective network where the heterogeneous transmission is reabsorbed into the edge structure. This amounts to rescaling edge probabilities:
\[
p_{ij}^{\text{SIR}} \, \beta = p_{ij}^{\text{\SMIR{}}} \, \beta_{P_jP_i} \quad \Rightarrow \quad p_{ij}^{\text{SIR}} = p_{ij}^{\text{\SMIR{}}} \frac{\beta_{P_jP_i}}{\beta}.
\]
That is, the SIR model on the effective graph has a homogeneous infection rate \(\beta\), while encoding heterogeneity within the connection structure.

\rev{Before showing some examples of this equivalence, let us add some remarks.
First, note that $\beta_{P_iP_j}$ factorizes into two node-specific terms, see Eq.~(\ref{eq:equivalence_0}). Therefore, the equivalence works properly when $p_{ij}^{\text{HeSIR}}$ can be expressed as the product of quantities per node, i.e. $p_{ij}^{\text{HeSIR}}=x_i x_j$. In this way, the node dependent term in $\beta_{P_iP_j}$ can be absorbed in the relative $x$ of $p_{ij}^{\text{SIR}}$. Although this restricts the class of admissible network models (excluding, for example, maximum-entropy ones~\cite{cimini2019statistical}), the most studied ones, as Erd\H{o}s--R\'enyi random graphs, Stochastic Block Models, Chung-Lu configuration models~\cite{chung2002connected}, Degree Corrected Block Models~\cite{barrer2011stochastic}, and Fitness models~\cite{caldarelli2002scale}, are included in this category.
Second, a natural question is what is the worth of such an equivalence. Besides the intellectual interest, the main point is that it is beneficial to translate non-trivial results from the SIR model into the HeSIR models. Therefore, theoretical results, as well as computational tools, developed in the SIR framework, can be safely used for the HeSIR, using this equivalence, with limited efforts. In this sense, consider that going, instead, from HeSIR to SIR is much more cumbersome, as when $\alpha_S\neq\alpha_I$, the equivalent contact network for the SIR model is a directed one, i.e. something non-physical, as standard contacts do not show a preferential direction. In a sense, a heterogeneity in the behaviour does not directly correspond in heterogeneous patterns of contact, but in something more involved.
In summary, while the mapping holds exactly only under specific structural assumptions, it provides a powerful conceptual bridge between parameter heterogeneity and structural heterogeneity.}

\subsection{Homogeneous networks: dSBM equivalence}

Consider the case where the interaction network of the \SMIR{} model is an ER graph with connection probability \(p_{ij}^{\text{\SMIR{}}} = p_\text{ER}\). \rev{The corresponding effective SIR network is defined by}%a directed SBM:
\[
p_{ij}^{\text{SIR}} = p_\text{ER}  [1 - (1 - \alpha_S)\delta_{P_i, H}]  [1 - (1 - \alpha_I)\delta_{P_j, H}],
\]
i.e., a directed SBM (\emph{dSBM}) where the edge probability depends on both the source and target group. When \(\alpha_I \neq \alpha_S\), the effective network is asymmetric: the probability of an edge from \(H\) to \(L\) differs from the reverse. Figure~\ref{fig:equiv_ER} shows the equivalence between the two models via a comparison of $R_\infty$, \rev{the frequency of infected individuals at the peak $I_\text{max}$, and the time of the peak $t_\text{max}$} across various values of $\beta/\gamma$.

\subsection{Heterogeneous networks: dDC-SBM equivalence}

When the \SMIR{} model is defined over a CM, such as a Chung-Lu network, the equivalent SIR model corresponds to a directed Degree Corrected Stochastic Block Model (\emph{dDC-SBM}). In this model, nodes have both in- and out-degree parameters that depend on their class, encoding differences in both susceptibility and infectivity.

\begin{figure*}[ht!]
    \centering
    \includegraphics[width=\textwidth]{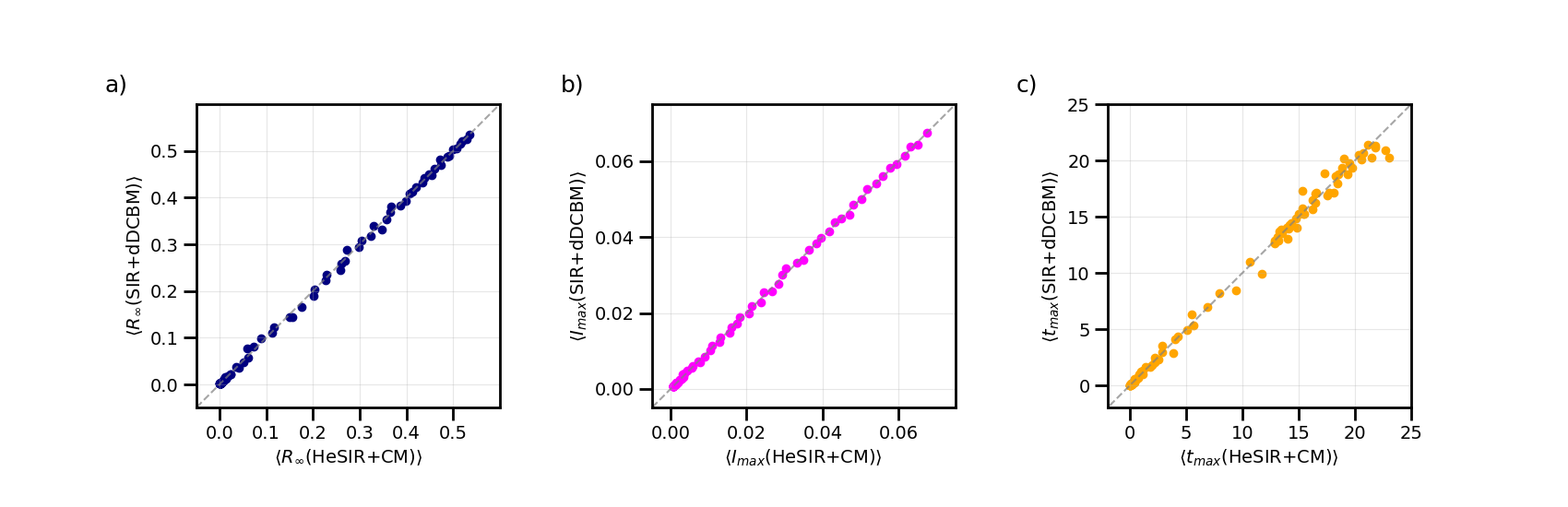}
    \caption{Model equivalence on heterogeneous networks. For $\beta/\gamma \in [0.01, 0.04]$, the average values of $R_\infty$ (panel a), \rev{fraction of infected individuals at the peak} $I_\text{max}$ (panel b), and \rev{the time of the peak} $t_\text{max}$ \rev{(panel c)} are shown for both the HeSIR and the equivalent SIR model, exhibiting perfect agreement. Averages are computed over 200 realizations. Parameters are set to $\alpha_I=2$, $\alpha_S=3$, and $p_H=0.4$. Node degrees in the CM network follow a Pareto distribution with exponent $a=2$.}
    \label{fig:equiv_CM}
\end{figure*}

Let \(k_i\) be the degree of node \(i\) in the CM.
We define effective in- and out-degrees as:
\[
k_i^{\text{in}} = f(P_i)  k_i \,, \quad
k_i^{\text{out}} = g(P_i)  k_i \,,
\]
\rev{where $f$  and $g$ are stretching functions that depend exclusively on the class of the node. Note that, for consistency,} \(f(P) = g(P) = 1\) if $\alpha_S=\alpha_I=1$.
Equating the transmission terms between \SMIR{} and SIR on the effective network leads to
\rev{
\begin{equation}\label{eq:equivalence_1}
\frac{\beta \, k_i k_j}{2m}  \frac{\beta_{P_jP_i} }{\beta} = \beta \, k_i^{\text{out}} k_j^{\text{in}}\frac{m_{P_iP_j}}{m_{P_i}^\text{out}m_{P_j}^\text{in}},  
\end{equation}
where $m=\sum_i k_i/2$ is the total number of links in the CM, $m_{P_iP_j}$, $m_{P_i}^\text{out}$ and $m_{P_j}^\text{in}$ are, respectively, the number of links from the block $P_i$ to the block $P_j$, the number of outgoing links from the block $P_i$ and the number of ingoing links in the block $P_j$, in the dDC-SBM.
Eq.~(\ref{eq:equivalence_1}) yields} the following set of constraints:
\begin{align*}
\frac{1}{2m} &= f(L) g(L)  \frac{m_{LL}}{m^{\text{out}}_{L} m^{\text{in}}_{L}}\,, \\
\frac{\alpha_S}{2m} &= f(H) g(L)  \frac{m_{LH}}{m^{\text{out}}_{L} m^{\text{in}}_{H}}\,, \\
\frac{\alpha_I}{2m} &= f(L) g(H) \frac{m_{HL}}{m^{\text{out}}_{H} m^{\text{in}}_{L}}\,, \\
\frac{\alpha_S \alpha_I}{2m} &= f(H) g(H) \frac{m_{HH}}{m^{\text{out}}_{H} m^{\text{in}}_{H}}\,.
\end{align*}
\rev{If \(\langle m_{L} \rangle\), \(\langle m_{H} \rangle\) are, respectively, the expected values under CM of the total number of links incident on $L$ and $H$, then
\[
m_P^{\text{in}} = f(P) \langle m_P \rangle, \quad m_P^{\text{out}} = g(P) \langle m_P \rangle.
\]
Therefore, the edge counts read:}
\begin{align*}
m_{LL} &= \frac{\langle m_{L} \rangle^2}{2m}\,    , \\
m_{LH} &= \alpha_S \frac{\langle m_{L} \rangle \langle m_{H} \rangle}{2m}\,, \\
m_{HL} &= \alpha_I \frac{\langle m_{L} \rangle \langle m_{H} \rangle}{2m}\,, \\
m_{HH} &= \alpha_S \alpha_I \frac{\langle m_{H} \rangle^2}{2m}\,.
\end{align*}
Solving for the stretching functions:
\[
\begin{cases}
\displaystyle \frac{f(H)}{f(L)} = \alpha_S, \quad f(L) = \frac{\langle m_{L} \rangle + \alpha_I \langle m_{H} \rangle}{2m}\,, \\
\displaystyle \frac{g(H)}{g(L)} = \alpha_I, \quad g(L) = \frac{\langle m_{L} \rangle + \alpha_S \langle m_{H} \rangle}{2m}\,.
\end{cases}
\]
Again, when \(\alpha_S \neq \alpha_I\), the resulting dDC-SBM is asymmetric, reflecting asymmetric transmission dynamics. The equivalence shows that the impact of behavioral heterogeneity in the \SMIR{} model can be fully encoded in the topology of an effective network for SIR dynamics. Figure~\ref{fig:equiv_CM} shows the equivalence between the two models via a comparison of $R_\infty$, $I_\text{max}$, and $t_\text{max}$ across various values of $\beta/\gamma$.

\section{Conclusion}

This work introduces the \SMIR{} model that extends the classical SIR model to incorporate heterogeneity in both susceptibility and infectivity.
%via two scalar parameters, \(\alpha_S\) and \(\alpha_I\). 
By modeling these traits as bimodally distributed across the population, capturing, for instance, compliant versus noncompliant behavioral types, we uncover how behavioral asymmetries significantly affect epidemic dynamics.
We derive closed-form analytical expressions for the epidemic threshold under homogeneous and heterogeneous network topologies. These estimates provide insight into the interplay between \(\alpha_S\), \(\alpha_I\), population composition, and the network mixing structure. In particular, we demonstrate that the epidemic threshold depends not merely on the average transmissibility, but on a weighted effective quantity that reflects the joint distribution of susceptibility and infectivity across contacts. Our theoretical predictions are corroborated by extensive stochastic simulations, showing excellent agreement across a wide range of settings, including networks with degree heterogeneity and varying levels of homophily.
Our findings reveal that the presence of heterogeneity in both susceptibility and infectivity is necessary to observe nontrivial effects such as the emergence of a resurgence zone, where a second epidemic transition appears beyond the classical threshold.

Our findings reveal that the presence of heterogeneity in both susceptibility and infectivity is necessary to observe nontrivial effects such as the emergence of a resurgence zone.
\rev{When both $\alpha_S>1$ and $\alpha_I>1$, in fact, infections can initially decline yet subsequently grow if $\left(\frac{\beta}{\gamma}\right)_{I_{\textrm{eff}}}<\frac{\beta}{\gamma}<\left(\frac{\beta}{\gamma}\right)_{I}$.
Only in the presence of this dual heterogeneity the threshold obtained from the classical percolation problem does not correspond to the initial growth of infections within the mean-field approximation, but rather to the initial increase of the \textit{effective} infections $I_{\textrm{eff}}=I_L+\alpha_I I_H$.}
This phenomenon is further amplified in homophilic networks and, remarkably, can also arise in purely structural models such as degree-heterogeneous networks, even in the absence of behavioral heterogeneity. These results highlight the importance of individual-level behavior and network structure in shaping epidemic outcomes.

Beyond theoretical contributions, the model has clear implications for epidemic control. In particular, interventions targeting individuals with low compliance (e.g., through testing, communication, or isolation) may be essential to prevent epidemic resurgence, as these individuals disproportionately drive transmission under HeSIR-like dynamics. Conversely, in networks with strong degree heterogeneity, monitoring and protecting high-degree nodes remains a priority regardless of behavioral assumptions.
Finally, by mapping the \SMIR{} dynamics onto equivalent SIR models on appropriately constructed networks (e.g., directed SBM or DC-SBM), we unify behavioral and structural heterogeneity into a common framework. This equivalence provides a powerful tool for extending standard epidemic theory to more realistic scenarios.
Future work may explore more finely resolved trait distributions, time-dependent behavior, empirical calibration using contact or genomic data, and the extension of these ideas to other epidemic models or coupled spreading processes.

\section{Data availability}
The data supporting the findings of this article are available at \cite{zenodo_data}.

\section{Acknowledgements}
All authors acknowledge support from the project “CODE – Coupling Opinion Dynamics with Epidemics”, funded under PNRR Mission 4 "Education and Research" - Component C2 - Investment 1.1 - Next Generation EU "Fund for National Research Program and Projects of Significant National Interest" PRIN 2022 PNRR, grant code P2022AKRZ9, CUP B53D23026080001. S.M. also acknowledges support from the Agencia Estatal de Investigaci\'on and Fondo Europeo de Desarrollo Regional (FEDER, UE) under project APASOS (PID2021-122256NB-C22) and the Mar\'ia de Maeztu program, project CEX2021-001164-M, funded by the  MCIN/AEI/10.13039/501100011033.

\begin{widetext}

\appendix

% \documentclass[10pt]{article}
% \usepackage{amsmath,amssymb,amsthm}
% \usepackage[a4paper,margin=1in]{geometry}
% \usepackage{graphicx}
% \usepackage{hyperref}
% \usepackage{authblk}
% \usepackage{tabularx}
% \usepackage{booktabs}
% \usepackage[numbers]{natbib}
% \usepackage{lineno}
% \usepackage{mathtools}

% \newcommand{\FM}[1]{{\textcolor{DarkBlue}{[FabioM: #1]}}}
% \newcommand\FC[1]{\textcolor{violet}{FC: #1}}
% \newcommand\FS[1]{{\color{magenta}#1}}
% \newcommand\SM[1]{{\color{orange}#1}}

% \newcommand{\bidc}[1]{\mathrm{BiDC}_{#1}}
% \newcommand{\monodc}[1]{\mathrm{MonoDC}_{#1}}

% \title{Supplementary Information: Impact of behavioral heterogeneity on epidemic outcome and its mapping into effective network topologies}

% \author[1]{Supplement to the main manuscript}
% \affil[1]{This document provides technical derivations, extended analyses, and computational details referenced in the main text.}

% \date{}

% \linenumbers

% \begin{document}

% \maketitle

\section{Epidemic threshold calculation}

As commonly done in the literature, the epidemic threshold of the HeSIR model can be determined based on bond percolation.
The activation probability $\phi_{ij}$ for an edge connecting group $i$ to group $j$ is:
\[
\phi_{LL} = 1 - e^{- \beta / \gamma},\;
\phi_{LH} = 1 - e^{-\alpha_I  \beta / \gamma},\; 
\phi_{HL} = 1 - e^{-\alpha_S  \beta / \gamma},\;
\phi_{HH} = 1 - e^{-\alpha_S \alpha_I  \beta / \gamma}.
\]
If $x_i$ denotes the probability that a vertex of type $i$ is \textit{not} connected to the giant component through a randomly chosen edge, we can formulate the fixed point system of equations
%\begin{widetext}
\begin{equation}
    \begin{cases}
        x_L = & \pi_{LL}(1 - \phi_{LL} + \phi_{LL} g_1(x_L)) + \pi_{LH}(1 - \phi_{LH} + \phi_{LH} g_1(x_H))\\
        x_H = & \pi_{HL}(1 - \phi_{HL} + \phi_{HL} g_1(x_L)) + \pi_{HH}(1 - \phi_{HH} + \phi_{HH} g_1(x_H))
    \end{cases}
    \label{eq:percolation_heterogeneous_app}
\end{equation}
where $g_1(x) \coloneqq \sum_k q_k x^k$ denotes the generating function of the network's excess degree distribution $q_k$.
System \eqref{eq:percolation_heterogeneous_app} always admits the trivial solution $(x_L,x_H)=(1,1)$, corresponding to the absence of a giant percolation cluster.
We look for conditions for a second solution to exist in $[0,1)\times [0,1)$.

System \eqref{eq:percolation_heterogeneous} can be rewritten as
\begin{equation}
\begin{cases}
    x_H & = f_H(x_L)\\
    x_L &= f_L(x_H)
\end{cases}
\label{eq:rewritten_system}
\end{equation}
where
\[
f_H(x_L) = g_1^{-1}\left(\frac{x_L-\pi_{LL}\left(1-\phi_{LL}+\phi_{LL}g_1(x_L)\right) -\pi_{LH}(1-\phi_{LH})}{\pi_{LH}\phi_{LH}}\right)
\] 
and 
\[
f_L(x_H) = g_1^{-1}\left(\frac{x_H-\pi_{HH}\left(1-\phi_{HH}+\phi_{HH}g_1(x_H)\right)-\pi_{HL}(1-\phi_{HL})}{\pi_{HL}\phi_{HL}}\right)
\]
\rev{After algebraic manipulation,} one sees that: 
% (i) $f_L(0)<0$; (ii) $f_H(0)<0$; (iii) $\ddot{f}_L\leq 0$ for all $x_H\in [0,1]$; (iv) $\ddot{f}_H\leq 0$ for all $x_L\in [0,1]$.
\[ f_L(0)<0,\; f_H(0)<0,\; \ddot{f}_L\leq 0 \text{ for all } x_H\in [0,1],\; \ddot{f}_H\leq 0 \text{ for all } x_L\in [0,1].
\]

We first consider the case in which $\dot{f}_L(1)< 0$, meaning that $f_L$ has a maximum at some $x_H^{\max}\leq 1$ with $f_L(x_H^{\max})> 1$.
In this case, $f_L=0$ at some $x_H^0>0$ and $f_L=1$ at some $x_H^1\in(x_H^0,1)$.
Now, let $h(x_H)=f_H(f_L(x_H))-x_H$.
$h$ is continuous, $h(x_H^0)=f_H(0)-x_H^0<0$ and $h(x_H^1)=f_H(1)-x_H^1>0$, so for the Intermediate Value Theorem there exists $\tilde{x}_H\in(x_H^0,x_H^1)$ such that $h(\tilde{x}_H)=0$, that is, $f_H(f_L(\tilde{x}_H))=\tilde{x}_H$.
If we call $\tilde{x}_L=f_L(\tilde{x}_H)$, we have $f_H(\tilde{x}_L)=\tilde{x}_H$.
In other words, $x_H=f_H(x_L)$ and $x_L=f_L(x_H)$ meet at $(x_L,x_H)=(\tilde{x}_L,\tilde{x}_H)\in [0,1)\times [0,1)$. 

Similarly, if $\dot{f}_H(1)< 0$, then $f_H$ has a maximum at some $x_L^{\max}\leq 1$ with $f_H(x_L^{\max})> 1$.
In this case, $f_H=0$ at some $x_L^0>0$ and $f_H=1$ at some $x_L^1\in(x_L^0,1)$.
Now, let $h(x_L)=f_L(f_H(x_L))-x_L$.
$h$ is continuous, $h(x_L^0)=f_L(0)-x_L^0<0$ and $h(x_L^1)=f_L(1)-x_L^1>0$, so for the IVT there exists $\tilde{x}_L\in(x_L^0,x_L^1)$ such that $h(\tilde{x}_L)=0$, that is, $f_L(f_H(\tilde{x}_L))=\tilde{x}_L$.
If we call $\tilde{x}_H=f_H(\tilde{x}_L)$, we have $f_L(\tilde{x}_H)=\tilde{x}_L$.
In other words, $x_L=f_L(x_H)$ and $x_H=f_H(x_L)$ meet at $(x_L,x_H)=(\tilde{x}_L,\tilde{x}_H)\in [0,1)\times [0,1)$. 

Finally, let us consider the case in which both $\dot{f}_L(1)\geq 0$ and $\dot{f}_H(1)\geq 0$.
We can invert $f_L$ in $[0,1]$ and rewrite system~\eqref{eq:rewritten_system} as
\begin{equation}
\begin{cases}
    x_H & = f_H(x_L)\\
    x_H &= f_L^{-1}(x_L)
\end{cases}
\label{eq:final_system}
\end{equation}
The system admits a solution in $[0,1)\times [0,1)$ if and only if $\dot{f}_L^{-1}(1)>\dot{f}_H(1)$ (with $\dot{f}_L^{-1}(1)=+\infty$ if $\dot{f}_L(1)=0$). In fact:
\begin{itemize}
\item On the one hand, $\dot{f}_L^{-1}(1)>\dot{f}_H(1)$ implies that some $\hat{x}_L$ sufficiently close to $1$ exists such that $f_H(\hat{x}_L)>f_L^{-1}(\hat{x}_L)$.
Now, let $l(x_L)=f_H(x_L)-f_L^{-1}(x_L)$.
Since $l(0)<0$ and $l(\hat{x}_L)>0$, for the IVT there exists $\tilde{x}_L\in(0,\hat{x}_L)$ such that $l(\tilde{x}_L)=0$, meaning that $f_H$ and $f_L^{-1}$ meet at $\tilde{x}_L<1$.
\item On the other hand, if $f_H$ and $f_L^{-1}$ meet at $\tilde{x}_L<1$, for the Mean Value Theorem there must exist $x'_L,x''_L\in[\tilde{x}_L,1]$ such that $\dot{f}_H(x'_L)=\dot{f}_L^{-1}(x''_L)$.
Since $\dot{f}_H$ is non-increasing ($\ddot{f}_H\leq 0$) and $\dot{f}_L^{-1}$ is non-decreasing ($\ddot{f}_L^{-1}\geq 0$), this implies $\dot{f}_H(1)< \dot{f}_L^{-1}(1)$.
\end{itemize}

In summary, recalling that $\dot{f}_L^{-1}(1) = \frac{1}{\dot{f}_L(1)}$, we expect to observe a giant percolation cluster if any of the following conditions holds: 
\begin{enumerate}
    \item $\dot{f}_L(1)< 0$
    \item $\dot{f}_H(1)< 0$
    \item $\dot{f}_L(1)\geq 0$, $\dot{f}_H(1)\geq 0$ and $\dot{f}_L(1)\dot{f}_H(1)<1$
\end{enumerate}

It is easy to check that the last condition is the most restrictive, so the epidemic threshold is determined by the solution of
\begin{equation*}
    \dot{f}_L(1)\dot{f}_H(1)-1 = 0
\end{equation*}
which can be rewritten as
\begin{equation}\label{eq:thr}
    \left(\pi_{LH}\pi_{HL}\phi_{LH}\phi_{HL}-\pi_{HH}\pi_{LL}\phi_{HH}\phi_{LL}\right) \dot{g}_1^2(1) + \left(\pi_{LL}\phi_{LL}+\pi_{HH}\phi_{HH}\right)\dot{g}_1(1) -1 = 0 
\end{equation}
If we assume $\beta / \gamma\ll 1$, we can linearize \eqref{eq:thr} with respect to $\beta / \gamma$ and ignore the first term of the equation.
Denoting $\alpha=\alpha_I\alpha_S$ and $\theta(p,h,\alpha)=\pi_{LL}+\alpha\pi_{HH}=(1 - h)(1 - p + \alpha p) + h(\alpha + 1)$, we obtain
\begin{equation}
   \left( \frac{\beta}{\gamma} \right)_c = \frac{1}{\left(\pi_{LL}+\alpha\pi_{HH}\right)\dot{g}_1(1)} = \frac{\langle k \rangle}{\langle k^2 \rangle - \langle k \rangle} \frac{1}{\theta(p,h,\alpha)}
\end{equation}
where the last equality follows from $\dot{g}_1(1)=\frac{\langle k^2 \rangle - \langle k \rangle}{\langle k \rangle}$.

\section{Comparison of epidemic thresholds with simulations}\label{sec:appendix_sim}

{In Figure \ref{fig:extra_sims_thresh} we present comparisons between numerical estimates of the epidemic threshold (using $V_{AR}$) and the theoretical predictions (equations \eqref{eq:threshold_true} and \eqref{eq:threshold_general} of the main text) across different combinations of \(\alpha_S\), \(\alpha_I\), \(p_H\), and \(\langle k\rangle\).
We further examine simulations for varying population sizes under homogeneous degree distributions (Figure \ref{fig:attackrates_N}) in order to assess finite-size effects.
Taken together, these analyses indicate that the findings remain consistent across all parameter configurations and population sizes examined.}

\begin{figure}[!htb]
    \centering
    \includegraphics[width=0.92\columnwidth]{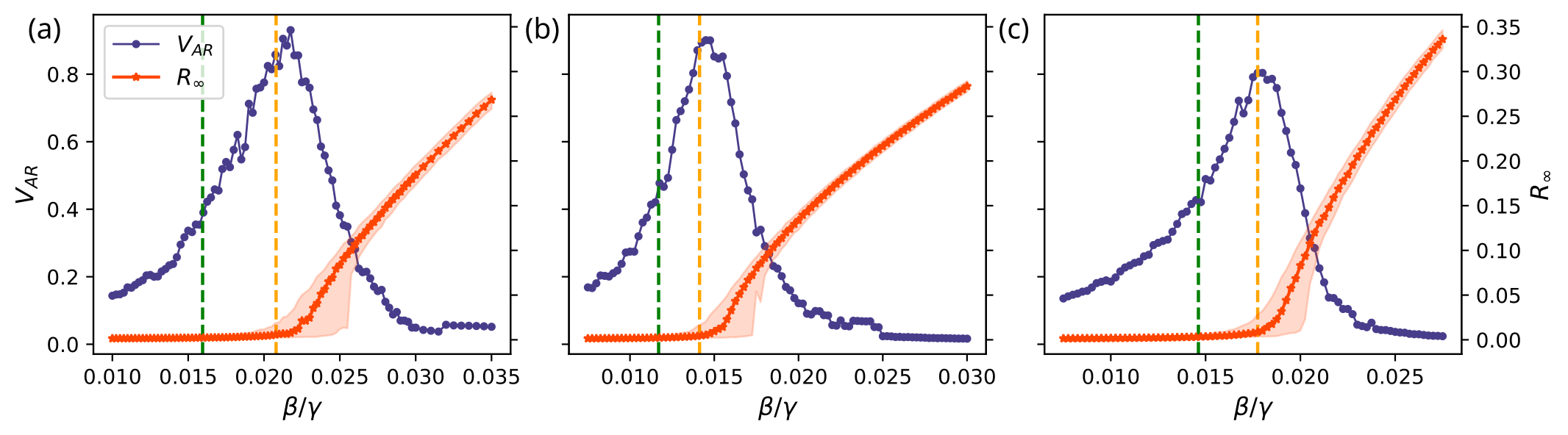}
    \includegraphics[width=0.92\columnwidth]{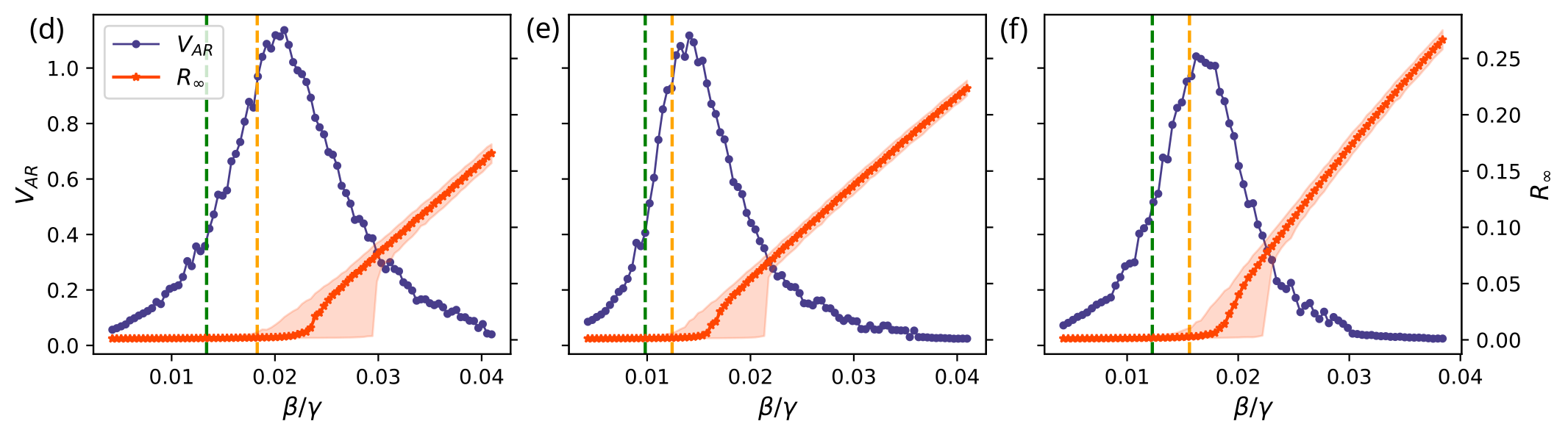}
    \caption{Epidemic threshold estimates for different configuration settings. 
    Yellow dashed lines: analytical threshold, computed numerically; Green dashed lines: approximation of the analytical threshold, as given by \eqref{eq:threshold_general}. The $R_\infty$ curves report the median value, with shaded areas showing the central 80\% of simulated outcomes. Results are obtained from 500 simulations per setting, {with networks of size $N=5\times 10^4$ and 50 seed nodes chosen uniformly at random. Results in the top panels (a-c) refer to a homogeneous degree network, with $h=0.5$, $\langle k \rangle=20$. The bottom panels (d-f), instead, refer to Pareto distributed degrees with exponent $a=2$, with $h=0.5$, $\langle k \rangle=12$. Each column explores a different combination of $\alpha_I$, $\alpha_S$ and $p_H$: $\alpha_I=2$, $\alpha_S=2$, $p_H=0.2$ in panels (a, d); $\alpha_I=2$, $\alpha_S=3$, $p_H=0.2$ in panels (b, e); $\alpha_I=2$, $\alpha_S=2$, $p_H=0.4$ in panels (c, f).}}
    \label{fig:extra_sims_thresh}
\end{figure}

\begin{figure}[ht!]
    {\centering
    \begin{subfigure}[b]{0.46\textwidth}
        \centering
        \includegraphics[width=\textwidth]{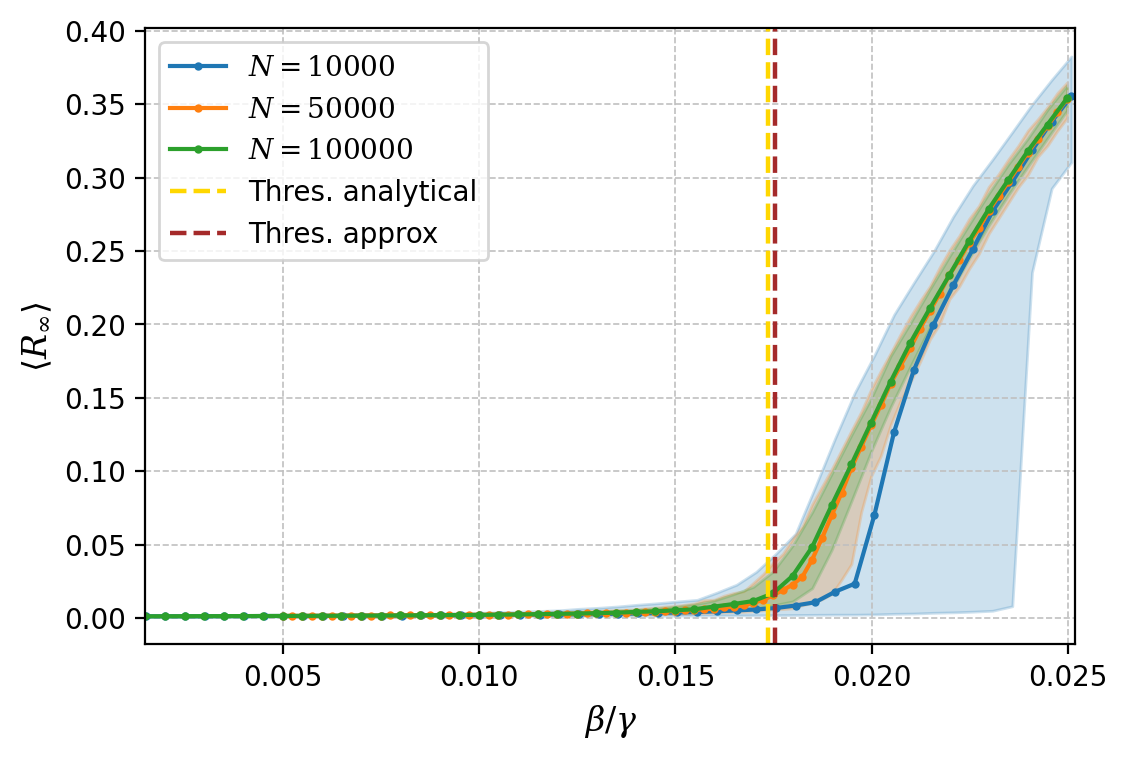}
        %{figures/Grafici_He_SIR/ev_ar_fabio/sbm_h_0_arvar_a_2.0_b_3.0_pm_0.4_fixedsw_Lut_c.png}
        \caption{$h=0$}
    \end{subfigure}
    \begin{subfigure}[b]{0.46\textwidth}
        \centering
        \includegraphics[width=\textwidth]{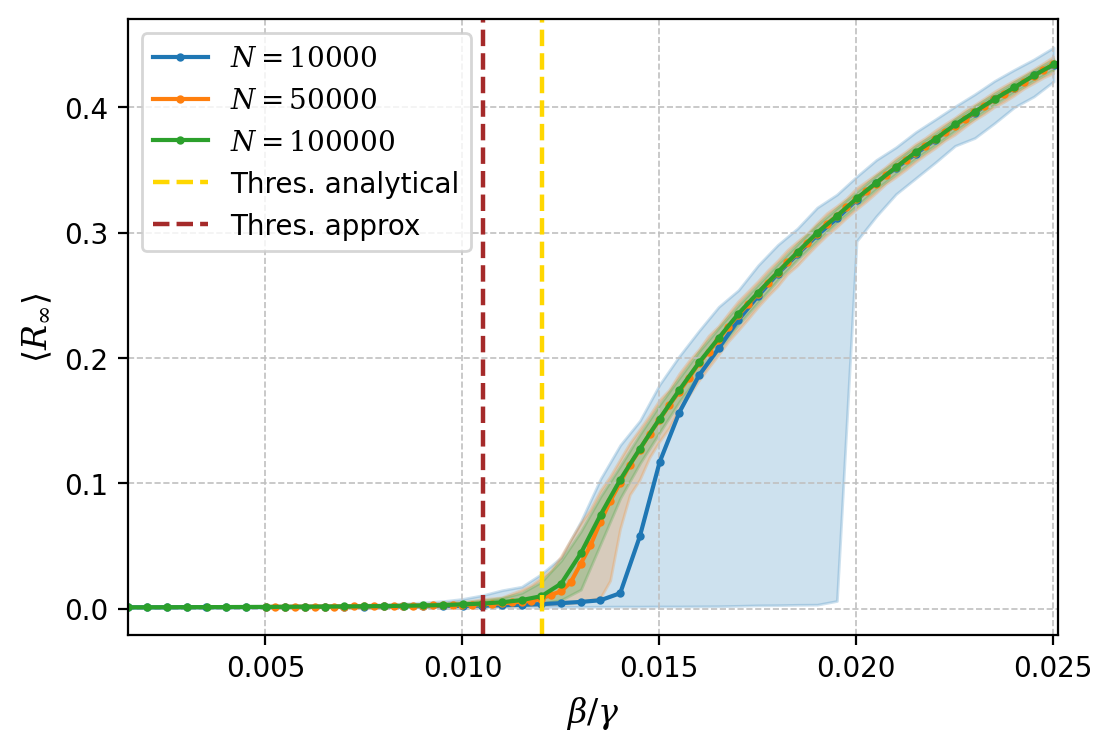}
        \caption{$h=0.5$}
    \end{subfigure}
    }
    \caption{{$R_\infty$ curves for different system sizes, for networks with homogenous degree distribution ($\langle k \rangle = 20$). The lines show the median value over 400 simulations, and the shaded areas show the central 80\% of simulated outcomes. In all simulations, $I(0)=10^{-3}$ and the seed nodes are chosen uniformly at random. Yellow dashed lines: analytical threshold, computed numerically; Brown dashed lines: approximation of the analytical threshold, as given by \eqref{eq:threshold_general}. Other parameters set as $\alpha_I=2$, $\alpha_S=3$, $p_H=0.4$.}}
    \label{fig:attackrates_N}
\end{figure}

% \bibliographystyle{plainnat}
% \bibliography{refs}  % Replace with your actual .bib file name

% \end{document}

\end{widetext}

% ---- OR ------- UncLHment this section to use bibtex
%\bibliographystyle{spmpsci}
\bibliography{refs}

@misc{cimini2019statistical,
    author = "Cimini, Giulio and Squartini, Tiziano and Saracco, Fabio and Garlaschelli, Diego and Gabrielli, Andrea and Caldarelli, Guido",
    archivePrefix = "arXiv",
    arxivId = "1810.05095",
    booktitle = "Nature Reviews Physics",
    doi = "10.1038/s42254-018-0002-6",
    eprint = "1810.05095",
    issn = "25225820",
    month = "jan",
    number = "1",
    pages = "58--71",
    publisher = "Nature Publishing Group",
    title = "{The statistical physics of real-world networks}",
    url = "http://www.nature.com/articles/s42254-018-0002-6",
    volume = "1",
    year = "2019"
}

@article{barrer2011stochastic,
  title = {Stochastic blockmodels and community structure in networks},
  author = {Karrer, Brian and Newman, M. E. J.},
  journal = {Phys. Rev. E},
  volume = {83},
  issue = {1},
  pages = {016107},
  numpages = {10},
  year = {2011},
  month = {Jan},
  publisher = {American Physical Society},
  doi = {10.1103/PhysRevE.83.016107},
  url = {https://link.aps.org/doi/10.1103/PhysRevE.83.016107}
}

@article{chung2002connected,
  title = {Connected Components in Random Graphs with Given Expected Degree Sequences},
  volume = {6},
  ISSN = {0218-0006},
  url = {http://dx.doi.org/10.1007/PL00012580},
  DOI = {10.1007/pl00012580},
  number = {2},
  journal = {Annals of Combinatorics},
  publisher = {Springer Science and Business Media LLC},
  author = {Chung,  Fan and Lu,  Linyuan},
  year = {2002},
  month = nov,
  pages = {125–145}
}

@article{caldarelli2002scale,
    author = "Caldarelli, G. and Capocci, A. and {De Los Rios}, P. and Mu{\~n}oz, M. A.",
    doi = "10.1103/PhysRevLett.89.258702",
    issn = "10797114",
    journal = "Physical Review Letters",
    month = "dec",
    number = "25",
    pages = "258702",
    pmid = "12484927",
    publisher = "American Physical Society",
    title = "{Scale-Free Networks from Varying Vertex Intrinsic Fitness}",
    url = "https://link.aps.org/doi/10.1103/PhysRevLett.89.258702",
    volume = "89",
    year = "2002"
}

@article{gibson_efficient_2000,
	title        = {Efficient Exact Stochastic Simulation of Chemical Systems with Many Species and Many Channels},
	author       = {Gibson, Michael A. and Bruck, Jehoshua},
	year         = 2000,
	journal      = {The Journal of Physical Chemistry A},
	volume       = 104,
	number       = 9,
	pages        = {1876--1889},
	doi          = {10.1021/jp993732q},
	issn         = {1089-5639},
	url          = {https://doi.org/10.1021/jp993732q},
	urldate      = {2024-09-16},
	shortjournal = {J. Phys. Chem. A},
	date         = {2000-03-01}
}

@article{karrer_stochastic_2011,
	title        = {Stochastic blockmodels and community structure in networks},
	author       = {Karrer, Brian and Newman, M. E. J.},
	year         = 2001,
	journal      = {Physical Review E},
	volume       = 83,
	number       = 1,
	pages        = {016107},
	doi          = {10.1103/PhysRevE.83.016107},
	url          = {https://link.aps.org/doi/10.1103/PhysRevE.83.016107},
	urldate      = {2024-06-28},
	shortjournal = {Phys. Rev. E},
	date         = {2011-01-21}
}

@article{beca-martinez_compliance_2022,
	title        = {Compliance with the main preventive measures of {COVID}-19 in {{Spa   in}}: The role of knowledge, attitudes, practices, and risk perception},
	shorttitle   = {Compliance with the main preventive measures of {COVID}-19 in Spain},
	author       = {Beca-Martínez, María Teresa and Romay-Barja, María and Falcón-Romero, María and Rodríguez-Blázquez, Carmen and Benito-Llanes, Agustín and Forjaz, María João},
	year         = 2022,
	journal      = {Transboundary and Emerging Diseases},
	volume       = 69,
	number       = 4,
	pages        = {e871--e882},
	doi          = {10.1111/tbed.14364},
	issn         = {1865-1682},
	url          = {https://onlinelibrary.wiley.com/doi/abs/10.1111/tbed.14364},
	urldate      = {2024-09-12},
	rights       = {© 2021 The Authors. Transboundary and Emerging Diseases published by Wiley-{VCH} {GmbH}},
	langid       = {english}
}

@article{deverna2024modeling,
	title        = {Modeling the amplification of epidemic spread by individuals exposed to misinformation on social media},
	author       = {DeVerna, Matthew R. and Pierri, Francesco and Ahn, Yong-Yeol and Fortunato, Santo and Flammini, Alessandro and Menczer, Filippo},
	year         = 2025,
	month        = apr,
	journal      = {npj Complexity},
	volume       = 2,
	number       = 1,
	pages        = {1--8},
	doi          = {10.1038/s44260-025-00038-y},
	issn         = {2731-8753},
	urldate      = {2025-05-13},
	copyright    = {2025 The Author(s)}
}

@article{volz2008sir,
	title        = {{{SIR}} dynamics in random networks with heterogeneous connectivity},
	author       = {Volz, Erik},
	year         = 2008,
	journal      = {Journal of mathematical biology},
	publisher    = {Springer},
	volume       = 56,
	pages        = {293--310}
}

@article{rodrigues2009heterogeneity,
	title        = {Heterogeneity in susceptibility to infection can explain high reinfection rates},
	author       = {Rodrigues, Paula and Margheri, Alessandro and Rebelo, Carlota and Gomes, M Gabriela M},
	year         = 2009,
	journal      = {Journal of theoretical biology},
	publisher    = {Elsevier},
	volume       = 259,
	number       = 2,
	pages        = {280--290}
}

@article{miller2007epidemic,
	title        = {Epidemic size and probability in populations with heterogeneous infectivity and susceptibility},
	author       = {Miller, Joel C.},
	year         = 2007,
	journal      = {Physical Review. E, Statistical, Nonlinear, and Soft Matter Physics},
	volume       = 76,
	number       = 1,
	pages        = {{010101}},
	doi          = {10.1103/PhysRevE.76.010101},
	issn         = {1539-3755},
	shortjournal = {Phys Rev E Stat Nonlin Soft Matter Phys},
	date         = {2007-07},
	pmid         = 17677396
}

@article{neri2011heterogeneity,
	title        = {Heterogeneity in susceptible--infected--removed ({{SIR}}) epidemics on lattices},
	author       = {Neri, Franco M and P{\'e}rez-Reche, Francisco J and Taraskin, Sergei N and Gilligan, Christopher A},
	year         = 2011,
	journal      = {Journal of The Royal Society Interface},
	publisher    = {The Royal Society},
	volume       = 8,
	number       = 55,
	pages        = {201--209}
}

@article{ming2016stochastic,
	title        = {Stochastic modelling of infectious diseases for heterogeneous populations},
	author       = {Ming, Rui-Xing and Liu, Jiming and Cheung, William KW and Wan, Xiang},
	year         = 2016,
	journal      = {Infectious diseases of poverty},
	publisher    = {Springer},
	volume       = 5,
	pages        = {1--11}
}

@article{bonaccorsi2016epidemics,
	title        = {Epidemics on networks with heterogeneous population and stochastic infection rates},
	author       = {Bonaccorsi, Stefano and Ottaviano, Stefania},
	year         = 2016,
	journal      = {Mathematical Biosciences},
	publisher    = {Elsevier},
	volume       = 279,
	pages        = {43--52}
}

@article{castellano2010thresholds,
	title        = {Thresholds for epidemic spreading in networks},
	author       = {Castellano, Claudio and Pastor-Satorras, Romualdo},
	year         = 2010,
	journal      = {Physical review letters},
	publisher    = {APS},
	volume       = 105,
	number       = 21,
	pages        = 218701
}

@article{holland1983stochastic,
	title        = {Stochastic blockmodels: First steps},
	author       = {Holland, Paul W and Laskey, Kathryn Blackmond and Leinhardt, Samuel},
	year         = 1983,
	journal      = {Social networks},
	publisher    = {Elsevier},
	volume       = 5,
	number       = 2,
	pages        = {109--137}
}

@article{newman2003structure,
	title        = {The structure and function of complex networks},
	author       = {Newman, Mark EJ},
	year         = 2003,
	journal      = {SIAM review},
	publisher    = {SIAM},
	volume       = 45,
	number       = 2,
	pages        = {167--256}
}

@article{pastor2015epidemic,
	title        = {Epidemic processes in complex networks},
	author       = {Pastor-Satorras, Romualdo and Castellano, Claudio and Van Mieghem, Piet and Vespignani, Alessandro},
	year         = 2015,
	journal      = {Reviews of modern physics},
	publisher    = {APS},
	volume       = 87,
	number       = 3,
	pages        = {925--979}
}

@article{pastor2001epidemic,
  title = {Epidemic Spreading in Scale-Free Networks},
  author = {Pastor-Satorras, Romualdo and Vespignani, Alessandro},
  journal = {Phys. Rev. Lett.},
  volume = {86},
  issue = {14},
  pages = {3200--3203},
  numpages = {0},
  year = {2001},
  month = {Apr},
  publisher = {American Physical Society},
  doi = {10.1103/PhysRevLett.86.3200},
  url = {https://link.aps.org/doi/10.1103/PhysRevLett.86.3200}
}

@book{barrat2008dynamical,
  title={Dynamical processes on complex networks},
  author={Barrat, Alain and Barthelemy, Marc and Vespignani, Alessandro},
  year={2008},
  publisher={Cambridge University Press},
  doi={10.1017/CBO9780511791383}
}

@book{newman2018networks,
  title={Networks},
  author={Newman, Mark},
  year={2018},
  edition={2nd},
  publisher={Oxford University Press},
  isbn={9780198805090}
}

@article{gou2017parameters,
title = {How heterogeneous susceptibility and recovery rates affect the spread of epidemics on networks},
journal = {Infectious Disease Modelling},
volume = {2},
number = {3},
pages = {353-367},
year = {2017},
issn = {2468-0427},
doi = {https://doi.org/10.1016/j.idm.2017.07.001},
url = {https://www.sciencedirect.com/science/article/pii/S246804271730009X},
author = {Wei Gou and Zhen Jin}
}

@article{clancy2014parameters,
title = {SIR epidemic models with general infectious period distribution},
journal = {Statistics \& Probability Letters},
volume = {85},
pages = {1-5},
year = {2014},
issn = {0167-7152},
doi = {https://doi.org/10.1016/j.spl.2013.10.017},
url = {https://www.sciencedirect.com/science/article/pii/S016771521300360X},
author = {Damian Clancy}
}

@article{DiDomenico2021,
	author = {Di Domenico, Laura and Sabbatini, Chiara E. and Bo{\"e}lle, Pierre-Yves and Poletto, Chiara and Cr{\'e}pey, Pascal and Paireau, Juliette and Cauchemez, Simon and Beck, Fran{\c c}ois and Noel, Harold and L{\'e}vy-Bruhl, Daniel and Colizza, Vittoria},
	date = {2021/12/06},
	doi = {10.1038/s43856-021-00057-5},
	id = {Di Domenico2021},
	isbn = {2730-664X},
	journal = {Communications Medicine},
	number = {1},
	pages = {57},
	title = {Adherence and sustainability of interventions informing optimal control against the COVID-19 pandemic},
	url = {https://doi.org/10.1038/s43856-021-00057-5},
	volume = {1},
	year = {2021}}

@article{Lloyd-Smith2005,
	author = {Lloyd-Smith, J. O. and Schreiber, S. J. and Kopp, P. E. and Getz, W. M.},
	date = {2005/11/01},
	doi = {10.1038/nature04153},
	id = {Lloyd-Smith2005},
	isbn = {1476-4687},
	journal = {Nature},
	number = {7066},
	pages = {355--359},
	title = {Superspreading and the effect of individual variation on disease emergence},
	url = {https://doi.org/10.1038/nature04153},
	volume = {438},
	year = {2005}}

@article{DeMeijere2021,
  title = {Effect of delayed awareness and fatigue on the efficacy of self-isolation in epidemic control},
  author = {De Meijere, Giulia and Colizza, Vittoria and Valdano, Eugenio and Castellano, Claudio},
  journal = {Phys. Rev. E},
  volume = {104},
  issue = {4},
  pages = {044316},
  numpages = {8},
  year = {2021},
  month = {Oct},
  publisher = {American Physical Society},
  doi = {10.1103/PhysRevE.104.044316},
  url = {https://link.aps.org/doi/10.1103/PhysRevE.104.044316}
}

@article{newman2002,
  title = {Spread of epidemic disease on networks},
  author = {Newman, M. E. J.},
  journal = {Phys. Rev. E},
  volume = {66},
  issue = {1},
  pages = {016128},
  numpages = {11},
  year = {2002},
  month = {Jul},
  publisher = {American Physical Society},
  doi = {10.1103/PhysRevE.66.016128},
  url = {https://link.aps.org/doi/10.1103/PhysRevE.66.016128}
}

@ARTICLE{Valganon2023-2,
  
AUTHOR={Valgañón, Pablo  and Lería, Unai  and Soriano-Paños, David  and Gómez-Gardeñes, Jesús },
         
TITLE={Socioeconomic determinants of stay-at-home policies during the first COVID-19 wave},
        
JOURNAL={Frontiers in Public Health},
        
VOLUME={11},

YEAR={2023},

URL={https://www.frontiersin.org/journals/public-health/articles/10.3389/fpubh.2023.1193100},

DOI={10.3389/fpubh.2023.1193100},

ISSN={2296-2565}}

@article{valganon2023,
	author = {Valga{\~n}{\'o}n, Pablo and Useche, Andr{\'e}s F. and Soriano-Pa{\~n}os, David and Ghoshal, Gourab and G{\'o}mez-Garde{\~n}es, Jes{\'u}s},
	date = {2023/09/30},
	doi = {10.1038/s41598-023-43685-8},
	id = {Valga{\~n}{\'o}n2023},
	isbn = {2045-2322},
	journal = {Scientific Reports},
	number = {1},
	pages = {16481},
	title = {Quantifying the heterogeneous impact of lockdown policies on different socioeconomic classes during the first COVID-19 wave in Colombia},
	url = {https://doi.org/10.1038/s41598-023-43685-8},
	volume = {13},
	year = {2023}}

@article{Lloyd2001,
author = {Lloyd, A. L. },
title = {Destabilization of epidemic models with the inclusion of realistic distributions of infectious periods},
journal = {Proceedings of the Royal Society of London. Series B: Biological Sciences},
volume = {268},
number = {1470},
pages = {985-993},
year = {2001},
doi = {10.1098/rspb.2001.1599},

URL = 
{https://royalsocietypublishing.org/doi/abs/10.1098/rspb.2001.1599},
eprint = {}
}

@article{Darbon2019,
author = {Darbon, Alexandre  and Colombi, Davide  and Valdano, Eugenio  and Savini, Lara  and Giovannini, Armando  and Colizza, Vittoria },
title = {Disease persistence on temporal contact networks accounting for heterogeneous infectious periods},
journal = {Royal Society Open Science},
volume = {6},
number = {1},
pages = {181404},
year = {2019},
doi = {10.1098/rsos.181404},

URL = {https://royalsocietypublishing.org/doi/abs/10.1098/rsos.181404},
eprint = {}
}

@article{krylova2013,
author = {Krylova, Olga  and Earn, David J. D. },
title = {Effects of the infectious period distribution on predicted transitions in childhood disease dynamics},
journal = {Journal of The Royal Society Interface},
volume = {10},
number = {84},
pages = {20130098},
year = {2013},
doi = {10.1098/rsif.2013.0098},

URL = {https://royalsocietypublishing.org/doi/abs/10.1098/rsif.2013.0098},
eprint = {}
}

@article{zhilan2000,
author = {Feng, Zhilan and Thieme, Horst R.},
title = {Endemic Models with Arbitrarily Distributed Periods of Infection I: Fundamental Properties of the Model},
journal = {SIAM Journal on Applied Mathematics},
volume = {61},
number = {3},
pages = {803-833},
year = {2000},
doi = {10.1137/S0036139998347834},

URL = { 
    
        https://doi.org/10.1137/S0036139998347834
    
    

},
eprint = { 
    
        https://doi.org/10.1137/S0036139998347834
    
    

}
}

@article{deArruda2020,
  title = {Impact of the distribution of recovery rates on disease spreading in complex networks},
  author = {de Arruda, Guilherme Ferraz and Petri, Giovanni and Rodrigues, Francisco A. and Moreno, Yamir},
  journal = {Phys. Rev. Res.},
  volume = {2},
  issue = {1},
  pages = {013046},
  numpages = {8},
  year = {2020},
  month = {Jan},
  publisher = {American Physical Society},
  doi = {10.1103/PhysRevResearch.2.013046},
  url = {https://link.aps.org/doi/10.1103/PhysRevResearch.2.013046}
}

@article{Cota2021,
doi = {10.1088/1367-2630/ac0c99},
url = {https://dx.doi.org/10.1088/1367-2630/ac0c99},
year = {2021},
month = {jul},
publisher = {IOP Publishing},
volume = {23},
number = {7},
pages = {073019},
author = {Cota, Wesley and Soriano-Paños, David and Arenas, A and Ferreira, Silvio C and Gómez-Gardeñes, Jesús},
title = {Infectious disease dynamics in metapopulations with heterogeneous transmission and recurrent mobility},
journal = {New Journal of Physics}
}

@article{montalban2022,
	author = {Montalb{\'a}n, Antonio and Corder, Rodrigo M. and Gomes, M. Gabriela M.},
	date = {2022/06/30},
	doi = {10.1007/s00285-022-01771-x},
	id = {Montalb{\'a}n2022},
	isbn = {1432-1416},
	journal = {Journal of Mathematical Biology},
	number = {1},
	pages = {2},
	title = {Herd immunity under individual variation and reinfection},
	url = {https://doi.org/10.1007/s00285-022-01771-x},
	volume = {85},
	year = {2022}
}

@inproceedings{mazza2024impact,
  title={The Impact of Heterogeneity on Epidemics: Insights from a Modified SIR Model},
  author={Mazza, Fabio and Colaiori, Francesca and Guarino, Stefano and Meloni, Sandro and Brambilla, Marco and Piccardi, Carlo and Pierri, Francesco and Saracco, Fabio},
  booktitle={International Conference on Complex Networks and Their Applications},
  pages={66--75},
  year={2024},
  organization={Springer}
}

@article{pastor2022advantage,
  title={The advantage of self-protecting interventions in mitigating epidemic circulation at the community level},
  author={Pastor-Satorras, Romualdo and Castellano, Claudio},
  journal={Scientific Reports},
  volume={12},
  number={1},
  pages={15950},
  year={2022},
  publisher={Nature Publishing Group UK London}
}

@article{clancy2018persistence,
  title={Persistence time of SIS infections in heterogeneous populations and networks},
  author={Clancy, Damian},
  journal={Journal of Mathematical Biology},
  volume={77},
  number={3},
  pages={545--570},
  year={2018},
  publisher={Springer}
}

@article{clancy2018precise,
  title={Precise estimates of persistence time for SIS infections in heterogeneous populations},
  author={Clancy, Damian},
  journal={Bulletin of Mathematical Biology},
  volume={80},
  number={11},
  pages={2871--2896},
  year={2018},
  publisher={Springer}
}

@dataset{zenodo_data,
  author       = {Mazza, Fabio and
                  Guarino, Stefano and
                  Saracco, Fabio},
  title        = {Impact of behavioral heterogeneity on epidemic
                   outcome and its mapping into effective network
                   topologies
                  },
  month        = dec,
  year         = 2025,
  publisher    = {Zenodo},
  doi          = {10.5281/zenodo.18017302},
  url          = {https://doi.org/10.5281/zenodo.18017302},
}

\end{document}